\documentclass{tlp}
\usepackage{ifthen}
\usepackage{url}
\usepackage{swipl}
\usepackage{amssymb}
\usepackage{upgreek}
\usepackage{color}
\usepackage{textcomp}

\newcommand\cplint{\texttt{cplint}}
\newcommand\cplintonswish{\cplint{} on SWISH}

\usepackage[pdftex]{graphicx}
\DeclareGraphicsExtensions{.pdf,.jpg,.png}
\graphicspath{{figs/}{./}}

\newcommand{\fnurl}[1]{\footnote{\url{#1}}}
\begin{document}
\bibliographystyle{acmtrans}

\title{Using SWISH to realise interactive web based tutorials for
       logic based languages}

\author[J. Wielemaker et al.]
{JAN WIELEMAKER \\
Centrum Wiskunde \& Informatica, Amsterdam, Netherlands \\
\email{J.Wielemaker@cwi.nl}
\and
 FABRIZIO RIGUZZI \\
\email{fabrizio.riguzzi@unife.it}
\and
BOB KOWALSKI \\
Imperial College, London\\
\email{r.kowalski@imperial.ac.uk}
\and
TORBJ\"ORN LAGER \\
University of Gothenburg\\
\email{lager@ling.gu.se}
\and
FARIBA SADRI \\
Imperial College, London\\
\email{fs@doc.ic.ac.uk}
\and
MIGUEL CALEJO \\
Logical Contracts\\
\email{mc@logicalcontracts.com}
}

\pagerange{\pageref{firstpage}--\pageref{lastpage}}
\volume{\textbf{?} (?):}
\setcounter{page}{1}

\maketitle
\begin{abstract}
Programming environments have evolved from purely text based to using
graphical user interfaces, and now we see a move towards web based
interfaces, such as Jupyter. Web based interfaces allow   for the
creation of
interactive documents that consist of text and programs, as well as
their output. The output can be rendered using web technology as, e.g.,
text, tables, charts or graphs. This approach is particularly suitable
for capturing data analysis workflows and creating interactive
educational material. This article describes SWISH, a web front-end for
Prolog that consists of a web server implemented in SWI-Prolog and a
client web application written in JavaScript. SWISH provides a web
server where multiple users can manipulate and run the same material,
and it can be adapted to support Prolog extensions. In this paper we
describe the architecture of SWISH, and describe two case studies of
extensions of Prolog, namely Probabilistic Logic Programming (PLP) and Logic
Production System (LPS), which have used SWISH to provide tutorial
sites.
\end{abstract}



\section{Introduction}
\label{sec:intro}

Interactive languages such as Prolog traditionally provide a
\jargon{console} to which one types commands (queries) and where the
output (answers) appear in textual form. The programming environment
consists of this interactive session, a code editor and optionally tools
for cross-referencing, profiling, testing, etc. These may be
integrated in a single front end called an \jargon{Integrated
Development Environment} (IDE) such as Emacs, Eclipse or Visual
Studio.

%
%
\jargon{Literate programming} combines text about a program with the
program itself in a single document. This field was pioneered by \citeN{knuth}, who created high quality documentation and a Pascal
program for \TeX{} from a single document. Note that in Knuth's vision the text was more design
documentation than merely simple comments. For program development
purposes literate programming evolved into generating program
documentation from source code based on properties of the code itself
such as class names or function names and \jargon{structured comments}.
Structured comments use the comment syntax of the target language,
avoiding the need for special programs to extract the program from the
document. Well known examples of this approach are the  JavaDoc and Doxygen.

Notably for domains where programs are used to perform calculations on
raw (experimental) data and produce derived data and charts, the
traditional literate programming approach has been extended to include
the result of the program such as a table or chart in the document.
The first interactive version of this approach, called \jargon{notebook
interfaces}, \jargon{computational notebooks} or \jargon{data science
notebooks} was introduced in 1988 by Mathematica 1.0 on the
Macintosh.\footnote{\url{https://en.wikipedia.org/wiki/Notebook_interface}}

Modern web technologies such as HTML5,
CSS3 and JavaScript allow for web based versions of notebooks,
simplifying their deployment. A good example of this is
Jupyter.\footnote{\url{http://jupyter.org/}} The notebook interface
paradigm is
claimed \cite{SWblog}
to be particularly suited to improve reproducibility in e.g., data
science workflows and to provide interactive educational material.

The benefits of interactive web based development environments and
notebooks for capturing workflows and education also apply to Prolog.
How should Prolog be part of these developments? There are two obvious
options: (1) provide support in an existing language independent
environment such as Jupyter or (2) develop a dedicated environment for
Prolog and related languages. Prolog has a number of distinctive
features that are hard to support in systems that are designed for
imperative object oriented languages. First, the program-is-data
feature, combined with the absence of keywords complicates syntax
highlighting, completion and explanations when \jargon{hovering}\footnote{Positioning the mouse
cursor over an area.},
which are found
in modern editors and expected by today's users. Second, non-determinism
allows the generation of multiple answers for one query, which requires a
different query/answer interface. Third, a large subset of queries is
side effect free, which allows multiple users to use a single server
through a web based interface. These observations motivated us to explore the
development of a pure Prolog oriented web based programming environment.

This article discusses SWISH,\footnote{\url{https://swish.swi-prolog.org}} an acronym for \jargon{SWI}-Prolog for
\jargon{SH}aring or \jargon{SWI}-Prolog \jargon{SH}ell. SWISH consists
of a web server written in SWI-Prolog and a client application written
in JavaScript. In this sense the design is similar to that of Jupyter.  SWISH
however is designed as a multi-user notebook that supports Prolog's
distinctive features and can be extended to facilitate extensions of
the Prolog language such as CHR, PLP and LPS.

This article describes the architecture of SWISH, how it is being
deployed to support Prolog education and how it has been extended to
support PLP and LPS.  The main contributions of this article are:

\begin{itemize}
    \item Provide a detailed description of a novel web-based Prolog
    environment for data analysis and education.

    \item Describe the potential of web-based Prolog.  Notably demonstrate
    how SWISH can be used to provide tutorial environments for derived
    languages.  The paper discusses two such environments:
    \cplint{} on SWISH (\secref{cplint}) for PLP and LPS on SWISH
    (\secref{lps-on-SWISH}) for LPS.

    \item Resolve technical challenges involved in running a web based
    IDE over low-bandwidth networks with high latency.  This notably
    poses challenges for server-assisted semantic syntax highlighting.
\end{itemize}

This article is organised as follows. After this introduction we provide
an overview of the functionality provided by SWISH in
\secref{application}.  In \secref{edu} we elaborate on education support.
\Secref{extending} discusses how SWISH can be used to provide a tutorial
site for two Prolog derived languages and is
followed by a description of the overall
architecture (\secref{arch}) and details about some of its critical
components. \secref{related} presents related work and
\secref{evaluation} discusses the impact, lessons learned and
portability.   We then conclude and present future directions.

\section{The SWISH application}
\label{sec:application}

\begin{figure}
    \includegraphics[width=\linewidth]{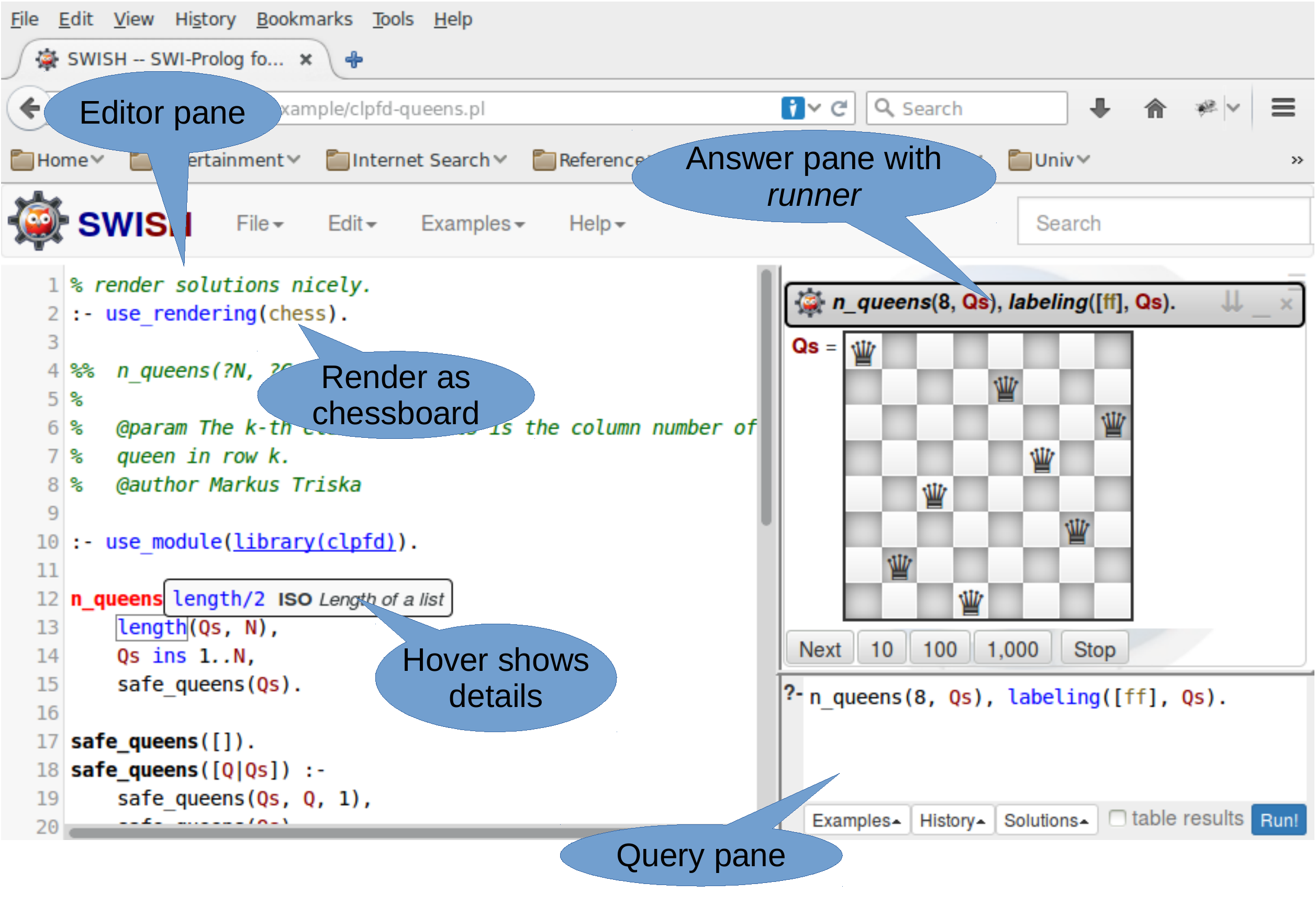}
    \caption{Screendump of SWISH in IDE mode. The left pane holds the
	     source code editor, while the top-right pane holds a query \jargon{runner}
	     that employs the current selected answer \jargon{renderer} and
	     buttons for continuing after the first answer. The bottom-right
	     pane holds the query editor with access to example queries stored
	     in the source, query history, solution modifiers, result
	     presentation and a \textsf{Run!} button to start the query.}
    \label{fig:swish}
\end{figure}

Before going into a technical description of the architecture in
\secref{arch} we present the functionality of the key components of
SWISH as it is exposed to the user. SWISH has two `faces'. The first is
an IDE `face' where it shows a program editor, a query editor and query
answers (see \figref{swish}). This is the original interface and it is
the most suitable for program development and testing. Programs created
in
this way may be saved and reused in other programs or notebooks using
the standard Prolog \index{include/1}\predref{include}{1} directive (\secref{gitty}). The second
`face' is the notebook interface described in \secref{notebook}. The
notebook interface is particularly suitable for educational applications
and recording data science workflows. First we describe the components
of the IDE interface. The notebook interface reuses these components.

\subsection{The program and query editor}
\label{sec:programeditor}
\label{sec:queryeditor}

The \jargon{program editor} uses
CodeMirror,\footnote{\url{https://codemirror.net/}} a JavaScript
text and code editor that we extended with
a Prolog mode. It  provides
semantic highlighting, template insertion and tooltips providing summary
information from the manual (\secref{codemirror}) and access to the
SWISH infrastructure for loading, storing and versioning of documents
(\secref{gitty}).

\begin{figure}
    \includegraphics[width=\linewidth]{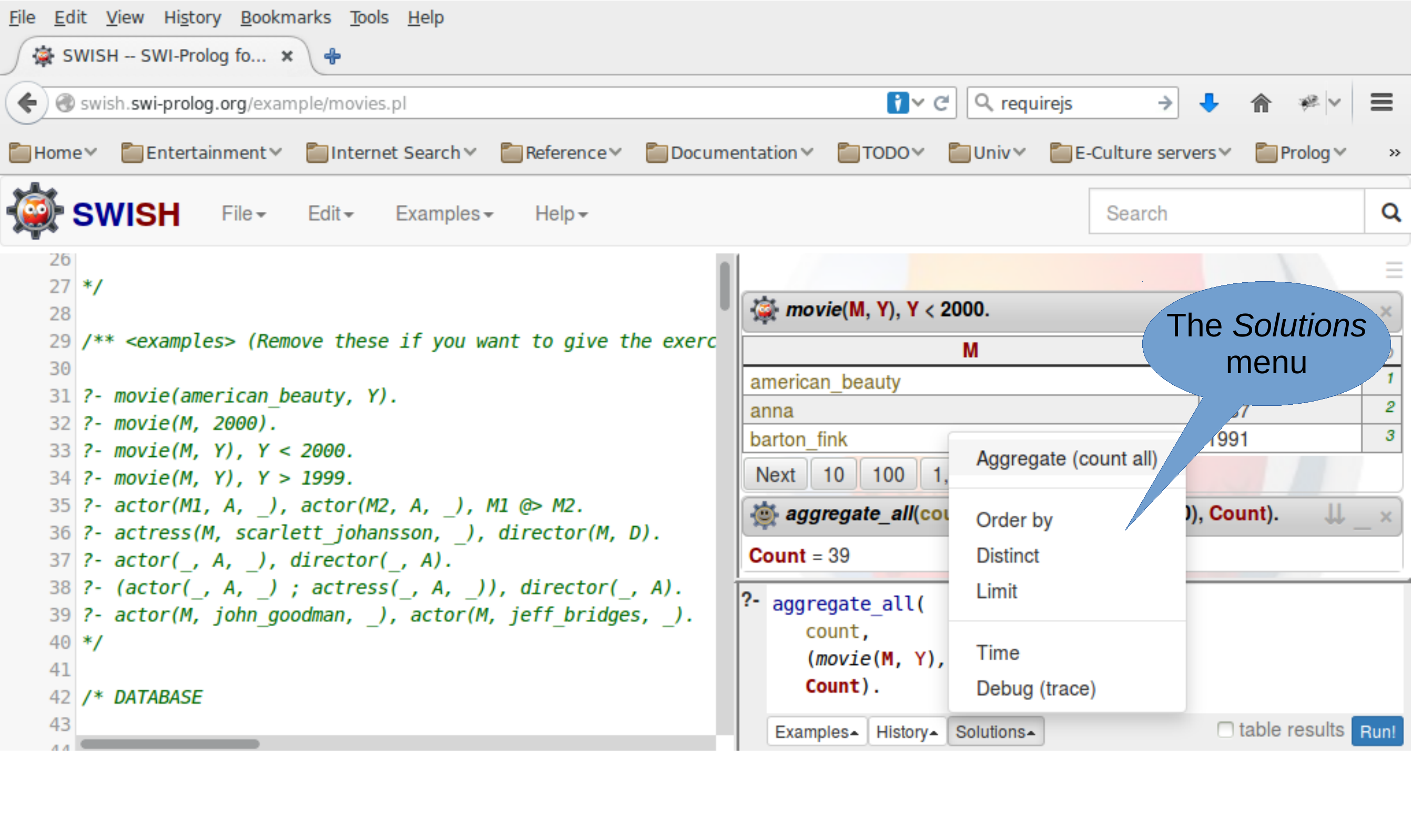}
    \caption{The \textsf{Solutions} menu can be used to count results,
	     order them, filter duplicates, etc. The upper runner shows
	     answers to the query as a table.}
    \label{fig:aggregate}
\end{figure}

The \jargon{query editor} is based on the same jQuery plugin that
realises the code editor and thus profits from syntax highlighting,
template insertion and tooltips. In addition, it provides three popup
menus:

\begin{description}
    \item[Examples] This menu is filled from structured comments
in the program.  Such a comment is formatted as below:

\begin{code}
/** <examples>
?- reverse([a,b], L).
?- reverse([a,b], [b,a]).
*/
\end{code}

\noindent
The examples menu provides a menu entry for adding the current query to
the program, either by creating a structured comment block as above or
adding the query to an existing block.

    \item[History] This menu provides access to previously executed queries.

    \item[Solutions]  This menu embeds an existing query in a meta-call to
alter the result. Currently provided operations are \textit{Aggregate
(count all)}, \textit{Order by}, \textit{Distinct}, \textit{Limit},
\textit{Time} and \textit{Debug (trace)}.  \Figref{aggregate} shows
how the menu is used to count the number of solutions of a goal.
\end{description}

SWISH allows the user to save the program in the code
editor. The user can assign the program a meaningful name or accept a
randomly created name. Once a program is saved, it is associated to an
URL of the form 
\begin{verbatim}
https://swish.swi-prolog.org/p/<name>
\end{verbatim}
that is
persistent: it can be used to retrieve the program later and to share it
with other users. Saved programs are versioned using technology inspired
by the GIT SCM,\footnote{\url{https://git-scm.com/}} where documents
(programs) and their meta-data (author, time, relation to previous
version etc.) are identified by a cryptographic hash of the data. This
approach guarantees integrity of the data and simplifies concurrent and
distributed management of versions. The versioning system allows for
using previous versions as well as examining changes. The user can log into the system: in this case
the identity of the user is associated with the saved program.

Each program is associated with a chat room where  users can
discuss the current program. SWISH has a general chat room for
discussions not related to a specific program.

After the user enters a query, the \textsf{Run!} button (bottom right hand corner) can be used to
execute the query. Pressing \textsf{Run!} collects the current program
and query and creates a \jargon{runner} instance which is discussed in
the next section.

\subsection{Running a query: runners in the answer pane}
\label{sec:answerpane}

The answer pane (top-right in \figref{swish}) is a placeholder for
\jargon{runners}. The answer pane provides a menu in the upper right corner for   operating on all
runners inside it. Possible actions are \textsf{Collapse all},
\textsf{Expand all}, \textsf{Stop all}, and \textsf{Clear}. Each runner
in the answer pane represents a query. The query may be completed,
running or waiting for user input. SWISH can manage multiple active
queries at the same time, up to an application defined maximum
(default~3). Once a runner has been created, its execution no longer
depends on the program and query editors.  While the query is executing
the user may prepare and start more queries.

Each runner provides its own set of controls to manage its query. During
execution,  a runner provides an \textsf{Abort} button. After query
evaluation completes with an answer and more answers may be available,
the runner allows the user to ask the next 1, 10, 100 or 1,000 results or to
\textsf{Stop} the query. In addition, the runner shows a \jargon{text
input} field if the application wants to read input and may show
debugger interaction buttons if the tracer is being used (see
\secref{tracer}).

A runner can render answers in two modes. The default classical Prolog
mode
renders the variable bindings as a sequence of \arg{Var}~=~\arg{Value}
statements. The \jargon{tabled}  mode displays a table where each column
represents a variable. The tabled mode is particularly appealing for
querying datasets (see \figref{aggregate}), while the default mode is more
suitable for rendering small amounts of complex answers such as the
chessboard position in \figref{swish}. By default, Prolog terms are
rendered as a nested HTML \elem{span} element where the rendered text is the
same as the output generated by Prolog's \index{writeq/1}\predref{writeq}{1} predicate.  The functor of a compound term
may be clicked to collapse the term to ellipses (\ldots).

\begin{figure}
    \includegraphics[width=\linewidth]{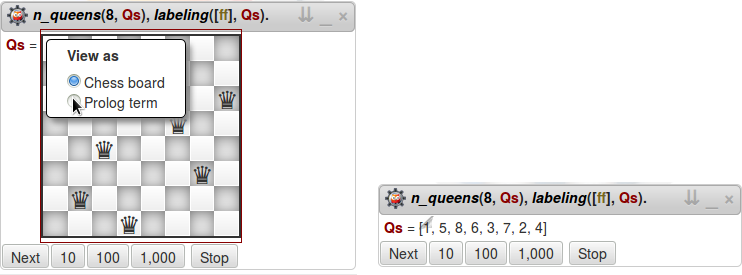}
    \caption{With the `chess' render library, a list of integers is
	     interpreted as queens on a chessboard. 
	     The numbers in the list are the indexes of the columns of the queens in each of the rows.
	      The user can
	     select rendering as a `Prolog term' to see the actual
	     term.}
    \label{fig:render}
\end{figure}

The server can provide \jargon{rendering libraries} that render Prolog
terms using dedicated HTML. In \figref{swish}, the `chess' renderer  translates a list of length $N$ holding integers in the
range $1\ldots N$ as a chessboard with queens. The interface allows for
switching the selected rendering as illustrated in \figref{render}.
The technical details are discussed in \secref{rendering}.

\subsubsection{Debugging}
\label{sec:tracer}

\begin{figure}
    \includegraphics[width=\linewidth]{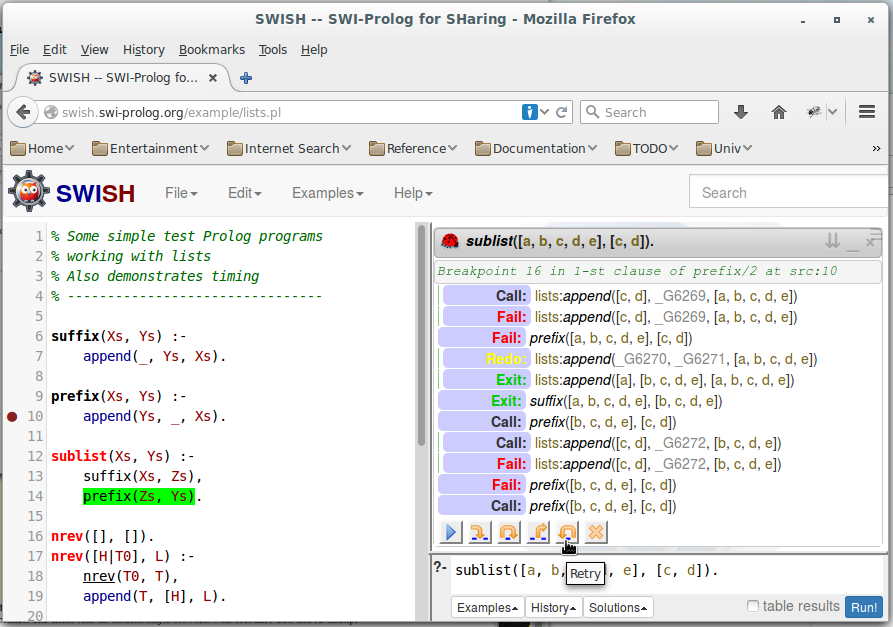}
    \caption{Debugging applications in SWISH}
    \label{fig:tracer}
\end{figure}

The SWISH debugger is based on the traditional 4-port debugging model
for Prolog \cite{Byrar}.
 The ports are events happening to goals in the procedural interpretation
of the program:
\begin{description}
\item[call] starting to prove the goal,
\item[exit] the goal was successfully proved,
\item[fail] the goal cannot be proved,
\item[redo]  backtracking for further solutions of the goal.
\end{description}
\Figref{tracer} shows the tracer in action on \index{sublist/2}\predref{sublist}{2} from
the \textit{Lists} example source. The debugger was triggered by a
break-point on line~10 set by clicking on the line-number in the code
editor. The debugging interaction is deliberately kept simple and
similar to traditional programming environments. A \jargon{retry} button
is added to the commonly seen `step into', `step over' and `step out'
for highlighting the feature of Prolog  of allowing the re-evaluation of a goal.

\subsection{The notebook interface}
\label{sec:notebook}

\begin{figure}
    \includegraphics[width=\linewidth]{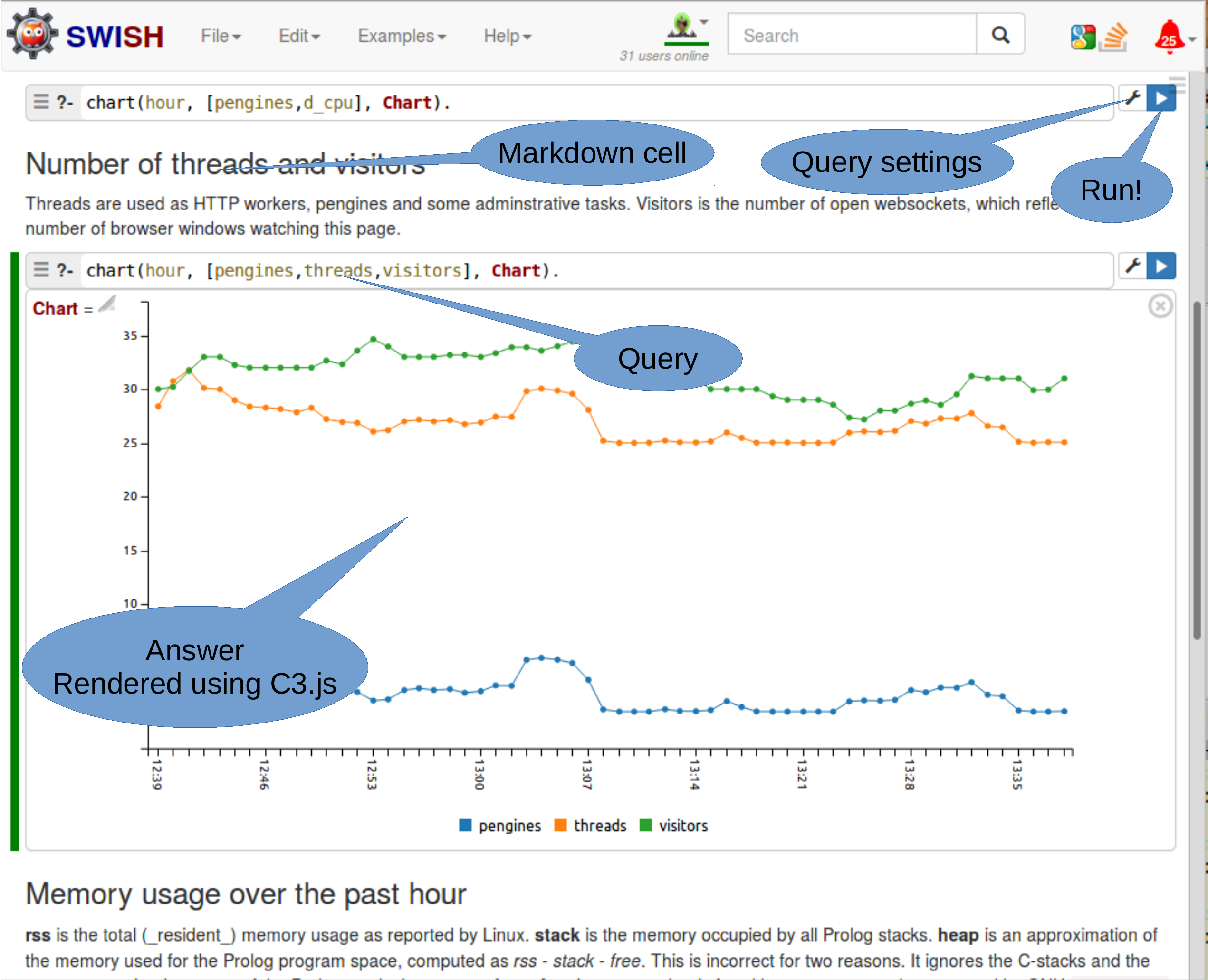}
    \caption{Screendump of SWISH in `notebook' mode.  The displayed
	     notebook is from the \textsf{Examples} menu, showing
	     statistics from the running SWISH server.  The visible
	     part contains two queries, the bottom one being executed.
	     The result is a chart showing the number of running
	     queries (Pengines), Prolog threads and visitors of the
	     website.  The result is rendered using C3.js.}
    \label{fig:notebook}
\end{figure}

A notebook is, inspired by Jupyter notebooks, a series of \jargon{cells}
that are displayed in page order (top-down), see Figure \ref{fig:notebook}. The notebook menu allows
cells to be added, copy/pasted, deleted or reordered. SWISH supports four cell types: \jargon{program}, \jargon{query},
\jargon{markdown} and \jargon{HTML}. The program and query cells are
instances of the program and query editors. The markdown and HTML
cells have two modes. By default they are displayed in rendered mode.
Double clicking them transforms them into a CodeMirror editor instance
with syntax highlighting. Clicking outside the cell re-enables
the rendered mode. Cells have a dynamic height that fits their content.
Program cells have a maximum height and may be collapsed to a single
line.

Notebooks have two types of programs, \jargon{local} and
\jargon{global}. A query that is being executed creates a
\jargon{runner} using the first local program cell above it and all
global program cells on the notebook. This is commonly used to realise
variants of a query. In this scenario,  global program cells are placed
(usually) at the bottom of the notebook and provide the shared code for
all queries. Just above the query cell we place a program cell that
completes the global program. For example, suppose we 
want to find paths between nodes in various graphs.
 In this case we write generic code for finding paths in graphs and
 we flag it as global. Then we may include in different local programs 
 the description of the different graphs in terms of their arcs: a query  below
 a local program can then perform pathfinding in the individual graph
 represented by the local program.

Notebook queries have a settings button that determines whether the
results are shown as a table or as  a traditional set of bindings, whether
or not the query should be executed when the page is loaded and how
many results to show initially.

\section{Using SWISH in Prolog education}
\label{sec:edu}
\label{sec:lpn}

The educational potential of SWISH was  obvious from its inception. SWISH
provides a  rich Prolog environment for basic educational
needs without installing software. The relatively simple and modern web
interface is appealing to students. The possibility of
saving programs assigning them an URL allows students to save their
assignment  and send the link to their teacher.

\begin{figure}
    \includegraphics[width=\linewidth]{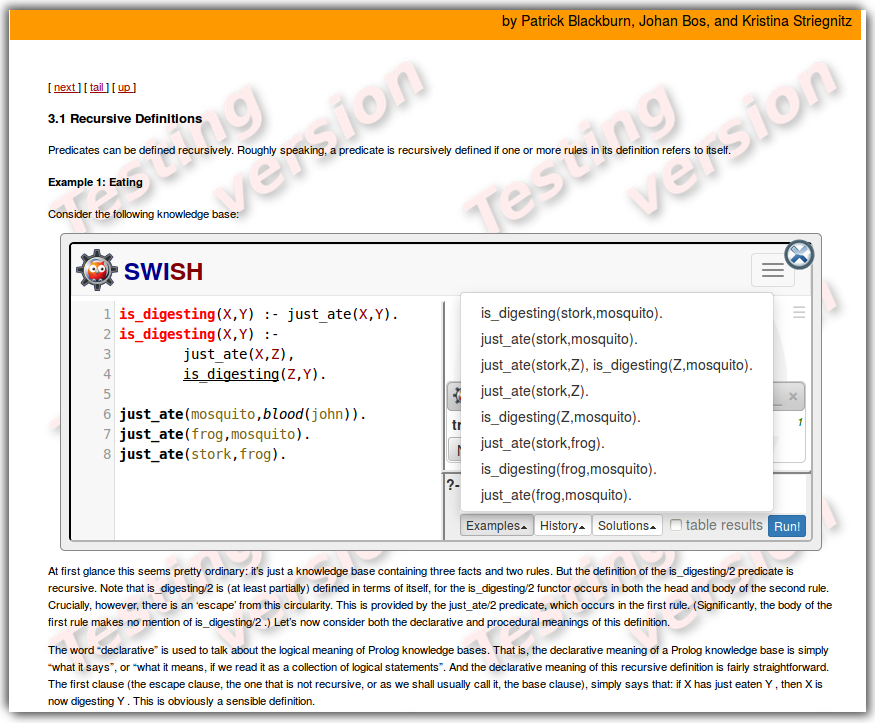}
    \caption{Screendump of Learn Prolog Now with an opened SWISH instance
	     that shows the collected source as well as example queries
	     from the following text that are classified as relating to
	     this source.  The embedded SWISH provides all functionality
	     available in the stand-alone SWISH. If the user presses the
	     close button, SWISH is removed and the original code
	     re-appears.}
    \label{fig:lpn}
\end{figure}

An obvious application is to exploit the ability to embed SWISH
using an \elem{iframe} element into existing online course material. The
first example was created by Jan Wielemaker for Learn Prolog
Now!\footnote{\url{http://www.learnprolognow.org}}, see \figref{lpn}.
Learn Prolog Now! is
an online version of a Prolog book by Patrick Blackburn, Johan Bos, and
Kristina Striegnitz \cite{blackburn2006learn}. We established a proof of
concept that embeds SWISH in the online course
material.\footnote{\url{http://lpn.swi-prolog.org}}
The current version rewrites the Learn Prolog Now! HTML on the fly, recognising source code and example queries. It is not yet good at recognising the relations between source code fragments and queries. Also Learn Prolog Now! needs some updating to be more compatible with SWI-Prolog. The
server served 304,205 pages (user pages, excluding javascript, css and
images) during 2016.

A recent example of an online book embedding SWISH is \textit{Simply Logical} by Peter
Flach.\footnote{\url{http://book.simply-logical.space/}}

From private mail exchange we know that several universities run local
copies of SWISH to support education. For example, at the University of
Ferrara SWISH is used to teach Prolog to the students of the Programming
Languages course. The students are given assignments in the form of
specifications and they have to write Prolog code implementing them. The
students then submit the assignment by sending the link of the saved
program or notebook to the teacher. The chatroom is one of the feature
most liked by the students: they use it mainly by exchanging messages
about the assignments among themselves. SWISH also considerably
simplifies laboratory management because it removes the need to install
and keep updated the Prolog stack on the laboratory machines. Moreover,
it makes Prolog accessible from the students' laptop straight away,
bypassing installation problems due to the wide variety of operating
systems and distributions in use by the students.

\cplintonswish{} is also used at the University of Ferrara for teaching
Statistical Relational Artificial Intelligence in the Data Mining \&
Analytics course: students are given an assignment whose solution is a
program in PLP, which they can develop in \cplintonswish{} and submit by
saving it in the system and sending the URL to the instructor. For this
course, the graphics capabilities of SWISH are really important, as the
output of probabilistic inference and learning can be best visualized
using graphs and diagrams. Therefore SWISH plays a very important role
in the dissemination and promotion of PLP in general and \cplint{} in
particular.

Finally, SWISH is also used in online education. ``Introduction to
SWI-Prolog'' is an online course taught by Anne Ogborn that run for 8
weeks from June 8th to August 6th 2018. It had 389 enrolled
participants. The course uses a variety of tools: the Moodle e-learning
platform, Youtube broadcasts, Slack channels, Twitter, forums and SWISH.
The instructors hands out weekly assignments, some of which are given as
SWISH notebooks with exercises that the student can solve by modifying
the notebooks and saving them with a different name. The saved notebooks
are tagged by the student with the \texttt{prologclass} tag so that the
instructor and fellow students can easily locate them. As of August 2nd
2018 the system includes 63 files tagged as \texttt{prologclass}.

\section{Extending SWISH}
\label{sec:extending}

SWISH can be extended to support languages built on top of Prolog. We
assume that the new language uses Prolog syntax, typically extended with
additional operators. We also assume we still have a \jargon{program}
and queries about this program that have zero, one or more answers. 
To build an
extension one needs to:

\begin{itemize}
    \item Provide syntax highlighting support for the target
          language.  This implies  using the SWI-Prolog library \texttt{prolog_colour.pl}.

    \item Provide additional \jargon{answer rendering} plugins.  See
          \secref{rendering}.

    \item Change the program or notebook \jargon{profiles}.  Profiles
	  are skeleton  or template programs and notebooks that are stored in a
	  directory \texttt{profiles}.  The user may create a new
	  program or notebook from one of these profiles.

    \item Add notebooks and programs that provide examples.  Examples
          are added to the directory \texttt{examples}.  The directory
	  contains a JSON index file that organises the content of the
	  \textsf{Examples} menu in the SWISH navigation bar.  Typically
	  only a few notebooks with a single markdown cell  describing
      the available examples are added to
	  the menu. 
\end{itemize}

\subsection{The \cplintonswish{} system}
\label{sec:cplint}

\cplintonswish{}\footnote{\url{http://cplint.ml.unife.it}}
\cite{RigBelLam16-SPE-IJ,AlbBelCot17-IA-IJ} is an
extension of SWISH for experimenting with
Probabilistic Logic Programming (PLP)
\cite{DBLP:journals/ml/RaedtK15},
the integration of logic programming with probability theory.
\cplintonswish{} provides a web interface to the \cplint{} system
\cite{RigSwi13-TPLP-IJ,Rig13-FI-IJ,BelRig13-IDA-IJ,BelRig15-TPLP-IJ}.

PLP is still a relatively new field and its user base is still growing.
One obstacle to its widespread adoption is the difficulty in the installation
of the various PLP systems. For example, \cplint{} is  distributed as
a SWI-Prolog
pack but its installation
requires the
compilation of an external C library together with
the CUDD library\footnote{\url{http://vlsi.colorado.edu/~fabio/}}
for handling Binary Decision Diagrams (BDDs). The compilation
is not automatic on some platforms, such as Windows, and requires
manual intervention. This may present a hurdle for some users.
\cplintonswish{} allows users to try \cplint{} without installing
anything on their machines, thus offering the opportunity   of
quickly getting a feeling of what PLP is and what \cplint{}
can do. Moreover, the possibility of saving, sharing and discussing
programs also
allows users to experiment with the system in a collaborative way,
which is particularly important when learning a new technology.

\cplintonswish{} handles PLP languages that follow the distribution semantics
\cite{DBLP:conf/iclp/Sato95ijar}: a probabilistic program defines a probability
distribution over normal programs called \jargon{instances},
and the probability of a query is the sum
of the probabilities of the instances where the query is true. This approach
has recently gained prominence  and it is adopted by many languages such as
Logic Programs with Annotated
Disjuctions (LPADs) \cite{VenVer04-ICLP04-IC},
ProbLog \cite{DBLP:conf/ijcai/RaedtKT07}
and PRISM \cite{DBLP:conf/iclp/Sato95ijar}.
Programs in any of these languages can be written
in \cplintonswish{}.

Each language offers the possibility
of expressing probabilistic choices: for example, LPADs allow disjunctions
in the head of clauses where each disjunct is annotated with a probability, the
meaning of a clause of this form being that, if the body is true, then one of
the head atoms is true with its corresponding probability.

In PLP the reasoning tasks are inference, parameter learning and structure
learning. In inference, the aim is to compute the probability of a query,
possibly given some evidence, from a program. Generating all possible instances
is not viable because there is an exponential number of them so smarter
algorithms must be adopted. One of the most successful approaches is
\jargon{knowledge compilation}, where the program is compiled into an
intermediate language from which the computation of the probability is easy.
\cplintonswish{} includes the PITA algorithm \cite{RigSwi13-TPLP-IJ}
that compiles the program to a BDD and computes the
probability with a dynamic programming algorithm. However, exact inference is
intractable in general so approximate algorithms have been developed.
\cplintonswish{} also includes MCINTYRE \cite{Rig13-FI-IJ}, which uses a Monte Carlo
approach: the algorithm repeatedly samples  the truth value of the query
and
the probability of the query is given by the relative frequency of the true
value.

In parameter learning, the user has a program for which he does not know
the probabilistic parameters and wants to induce them from data.
In structure learning, the user wants to induce both the structure of the
clauses and the parameters from data.
In \cplintonswish{} learning can be performed from
sets of interpretations, each expressing a possible state of the world.
Some predicates are identified as target, and true and false atoms of the target predicates
in the
interpretations are used as positive and negative examples respectively.

 EMBLEM \cite{BelRig13-IDA-IJ} performs parameter learning in \cplintonswish{} 
using a special dynamic programming algorithm operating on BDDs.
For structure learning, \cplintonswish{} includes SLIPCOVER
\cite{BelRig15-TPLP-IJ}, which performs a  search in the space of clauses
and scores them using the  likelihood of the data after parameter learning
by EMBLEM, and LEMUR \cite{DBLP:journals/ml/MauroBR15}, which is similar
to SLIPCOVER but uses Monte Carlo tree search.

\cplintonswish{} includes many examples of inference from various domains:
reasoning about actions,
random walks,
marketing,
natural language,
biology,
puzzles,
genetics,
model checking,
medicine,
games,
social networks,
filtering,
Bayesian estimation and
regression.
These examples show the various features of the inference modules:
encoding Markov Logic Networks and stochastic logic programs,
continuous random variables,
particle filtering,
likelihood weighting,
Metropolis-Hastings sampling,
rejection sampling,
argument sampling,
causal inference and
computation of expectations, see \cite{RigBelLam16-SPE-IJ,AlbBelCot17-IA-IJ,NguRig17-IMAKE-BC} for more details on the examples.

\begin{figure}
    \includegraphics[width=\linewidth]{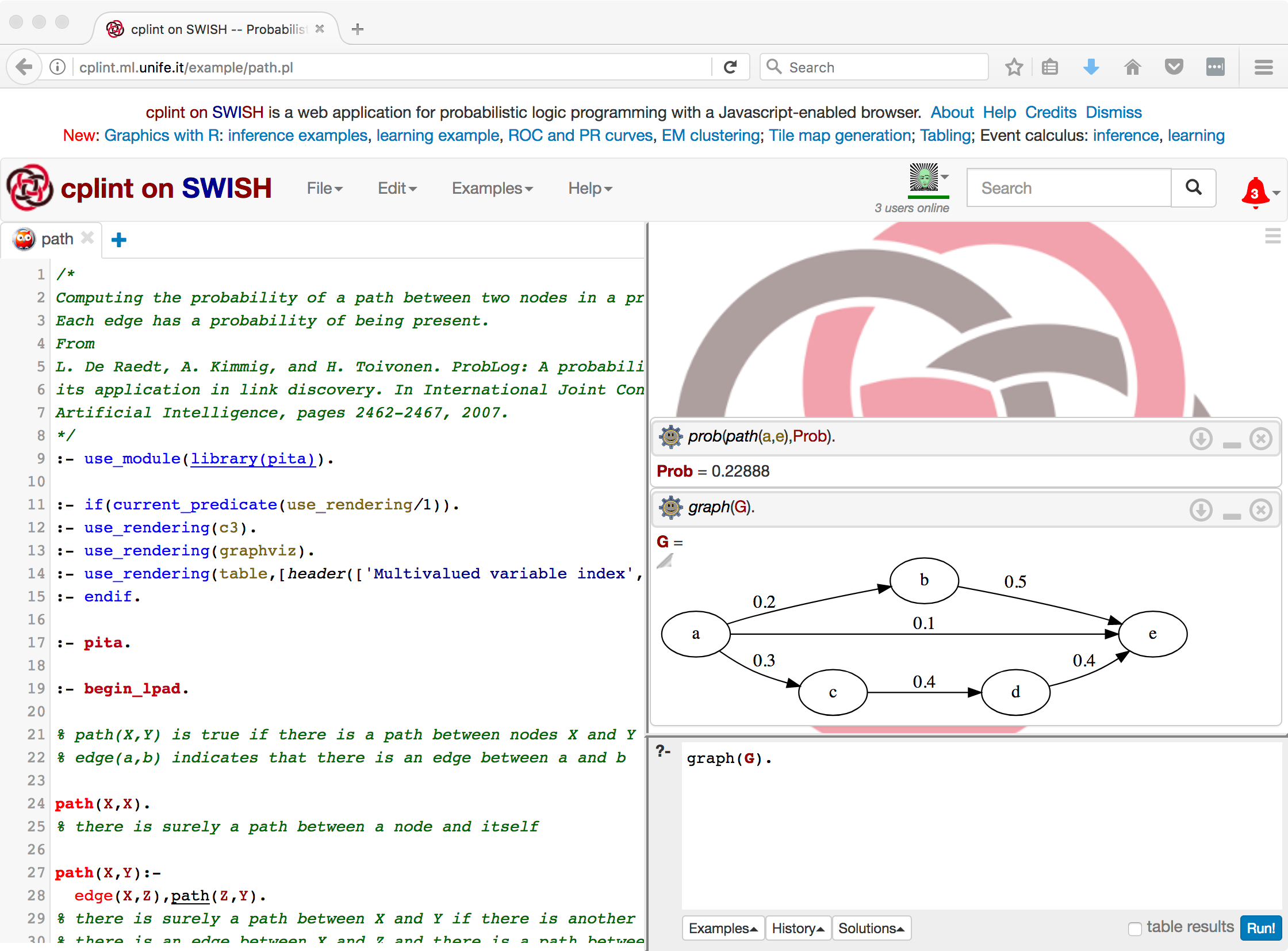}
    \caption{Screendump of \cplintonswish{} with opened
    probabilistic graph example.}
    \label{fig:cplint-path}
\end{figure}

\cplintonswish{} exploits the graphics capabilities of SWISH.
For example, Figure \ref{fig:cplint-path} shows a program
for the computation of the probability of the existence of a
path between two nodes in a probabilistic graph
(also known as the \jargon{network reliability problem}).
The graphviz rendering library
of SWISH is used to draw the probabilistic graph.
The same library can be used to draw the BDD that is built by
the PITA algorithm for answering a query.

\begin{figure}
    \includegraphics[width=\linewidth]{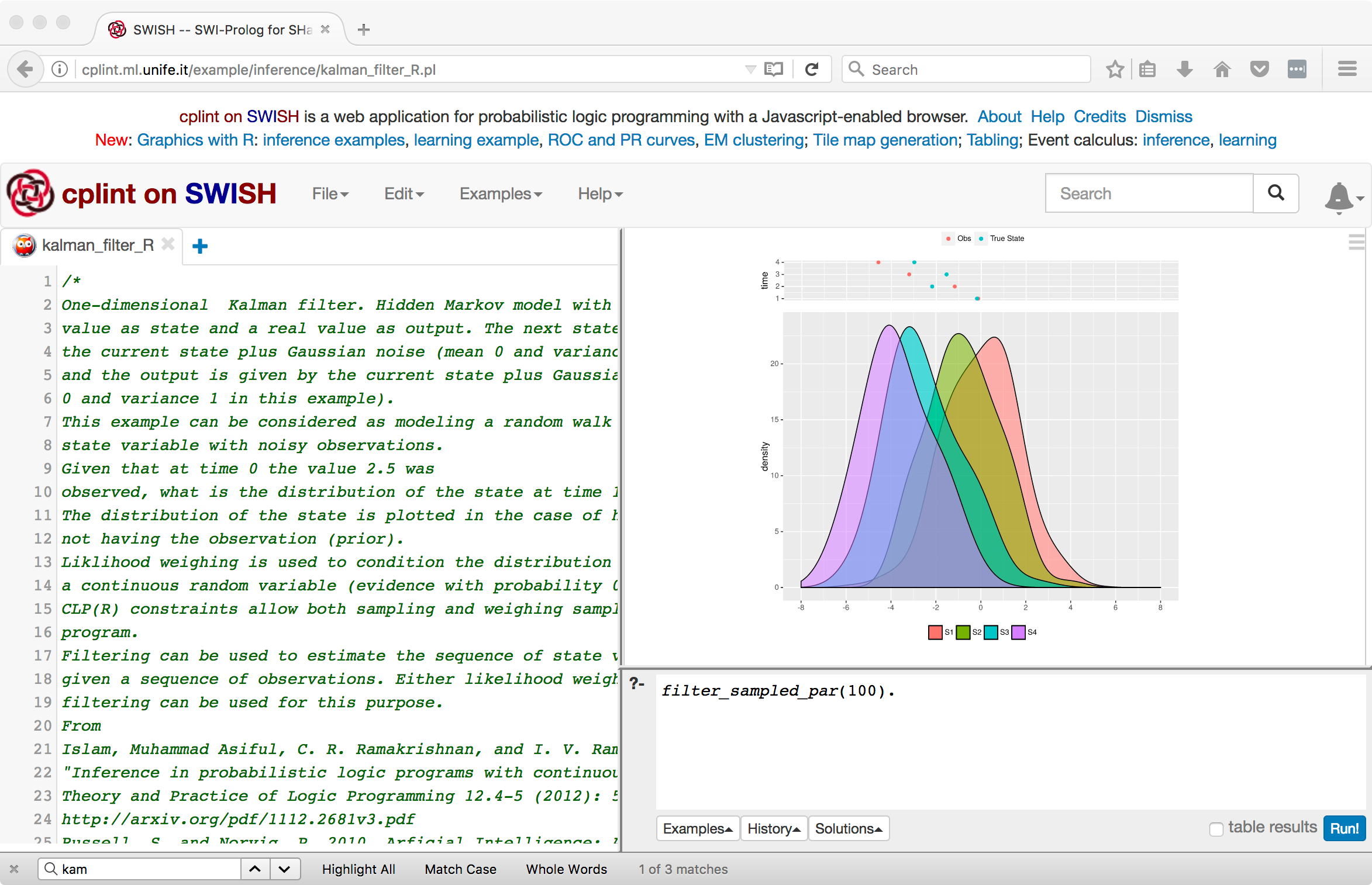}
    \caption{Screendump of \cplintonswish{} with opened
    Kalman filter example together with the results of
    a MCINTYRE query graphed with R.}
    \label{fig:cplint-kalman}
\end{figure}

\cplintonswish{} also uses the R for SWISH library,
which provides an interface to R both for computation and for
graphics. \Figref{cplint-kalman} presents
a one-dimensional Kalman filter example with a graph
that shows how the probability density on the position
of the target changes with time.

\begin{figure}
    \includegraphics[width=\linewidth]{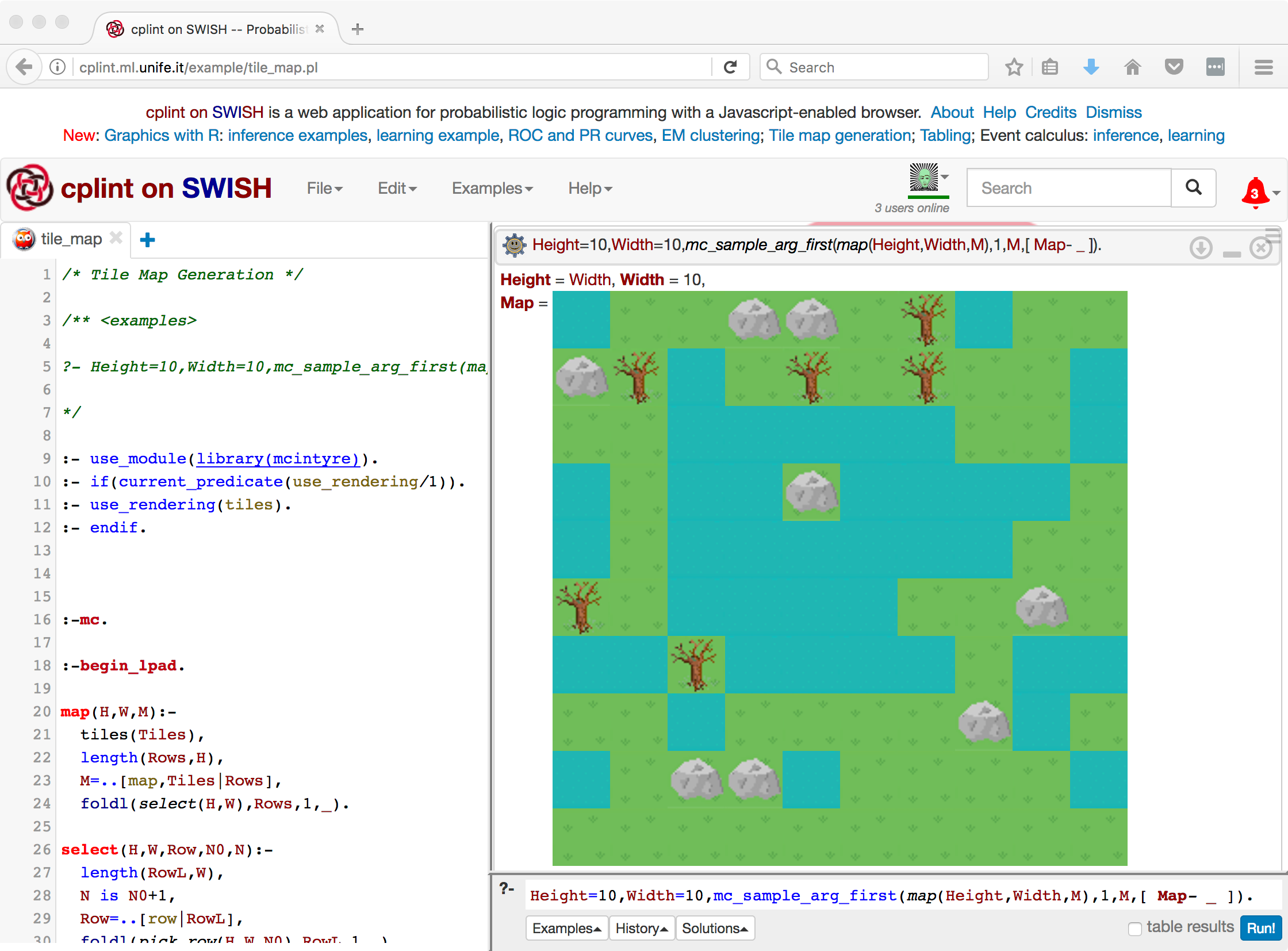}
    \caption{Screendump of \cplintonswish{} with opened
    tile map generation example.}
    \label{fig:cplint-map}
\end{figure}

SWISH also offers the possibility of adding rendering plugins.
\cplintonswish{} for example adds a plugin for drawing
two dimensional tile maps. \Figref{cplint-map}
shows an example for the random generation of a two-dimensional
tile map for a video game.

\cplintonswish{} also includes learning examples, both for
parameter learning and for structure learning with SLIPCOVER
and LEMUR: predicting  whether a machine should be fixed,
classifying  pictures containing geometrical objects
(Bongard problems),
predicting the shopping behaviour of people,
learning the parameters of hidden Markov models,
predicting whether a molecule is an active mutagenic agent
(the famous mutagenesis benchmark dataset  \cite{DBLP:journals/ai/SrinivasanMSK96}),
predicting the rating of courses of a university
and learning the probabilistic effects of actions in the Event Calculus.

\begin{figure}
    \includegraphics[width=\linewidth]{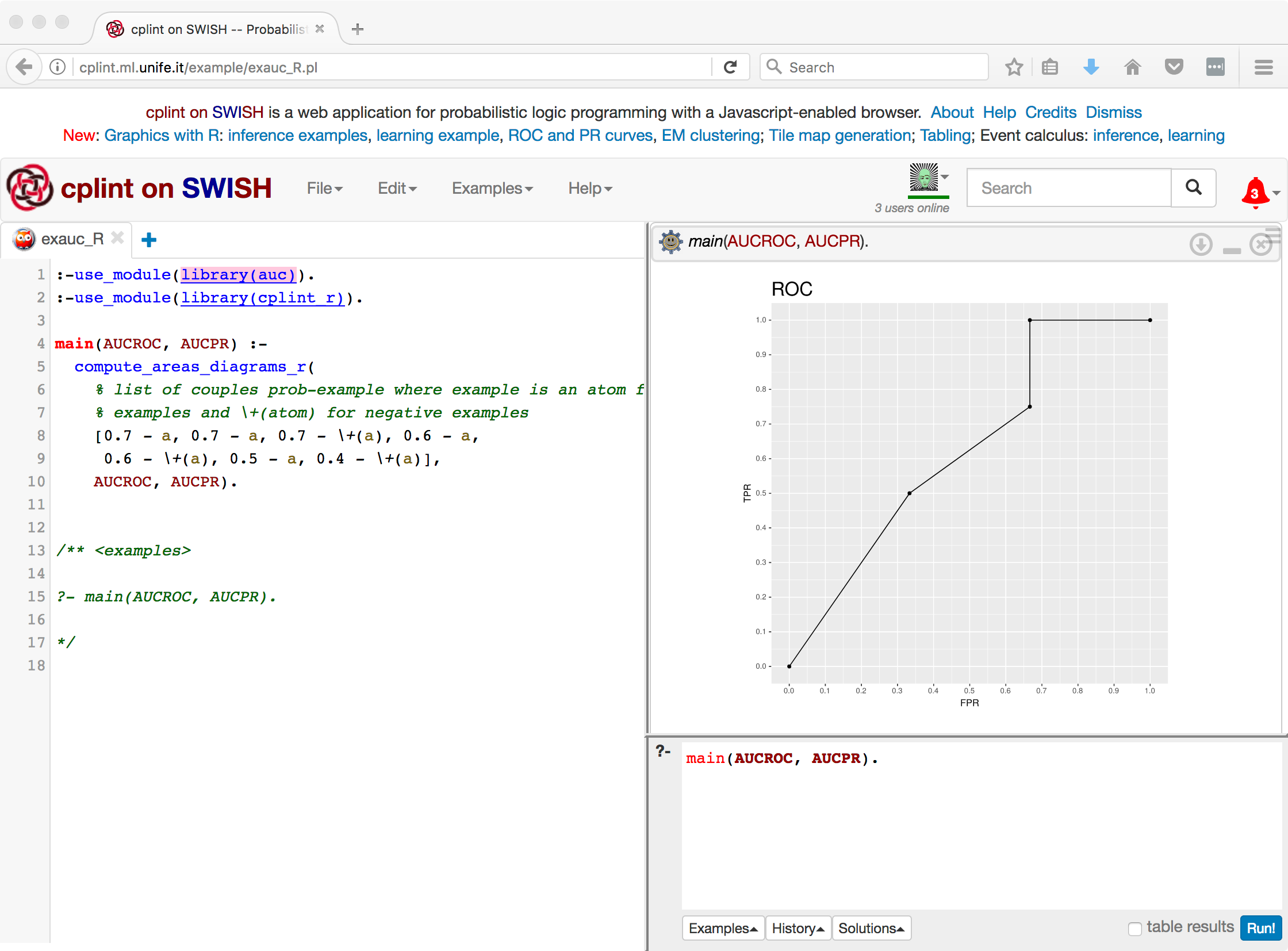}
    \caption{Screendump of \cplintonswish{} with opened
    ROC and PR curves example.}
    \label{fig:cplint-auc}
\end{figure}

The learned programs are shown in the answer pane using a specific
rendering library and can be applied to a testing set. Data
analysts  often find it useful to
draw the Received Operating Characteristic (ROC)
and Precision Recall (PR) curves of the performance of a model
on the test set and to compute
the areas under these curves. \cplintonswish{} includes
the \verb|auc|
SWI-Prolog pack for such purpose and can draw the curves using
either R or C3.js: for example, Figure \ref{fig:cplint-auc}
shows a ROC curve drawn with R.

\cplintonswish{} also includes Aleph \cite{aleph}, one of the most
influential Inductive Logic Programming (ILP) systems. It learns
normal logic programs from examples and background knowledge and
can be used as a baseline for probabilistic ILP systems.
Aleph has been ported from Yap to SWI-Prolog and transformed into a
module with conditional compilation directives as the \cplint{}
algorithms in order to work in SWISH. It is now available
as a SWI-Prolog pack.

The inclusion of these machine learning systems makes \cplintonswish{}
an example of the use of SWISH for data analysis.

\cplintonswish{} offers program and notebook profiles for each of
the inference and learning algorithms containing code skeletons
for using the algorithms. The syntax highlighter of SWISH was extended
to correctly
 emphasize the probabilistic constructs.

\subsection{The LPS on SWISH system}
\label{sec:lps-on-SWISH}


LPS (Logic-based Production System) \cite{lpsmodelgeneration,lpswithoutlp} is an extension of logic programming, which gives a logical interpretation to imperative sentences, by interpreting them as goals in first-order logic. Computation in LPS satisfies goals by generating a model that makes them true. It generates models for  \jargon{reactive rules} of the form \jargon{if antecedent then consequent}, by performing actions to  make  \jargon{consequents} true whenever  \jargon{antecedents} become true. 

Reactive rules in LPS behave like production rules in production systems and like active rules (or triggers) in active databases. But, unlike production rules and active rules, which have only an imperative reading, reactive rules in LPS have a combined declarative and imperative interpretation:

\begin{quote}
	To \textbf{make} \emph{if antecedent then consequent} true,
	
	\textbf{make} $consequent$ true whenever $antecedent$ becomes true.
\end{quote}
Models in LPS are defined by logic programs, which represent beliefs, and are generated incrementally, starting from an initial state, observing a stream of external events, executing a stream of actions, and updating the current state. Updates are performed using causal laws, which define the time-varying facts (or fluents) initiated and terminated by events, including both external events and actions. Causal laws are based on the Event Calculus \cite{eventcalculus}, but are implemented destructively, as in imperative computer languages. Analogously, events are processed as streams, without remembering their past.   

\begin{figure}
	\includegraphics[width=\linewidth]{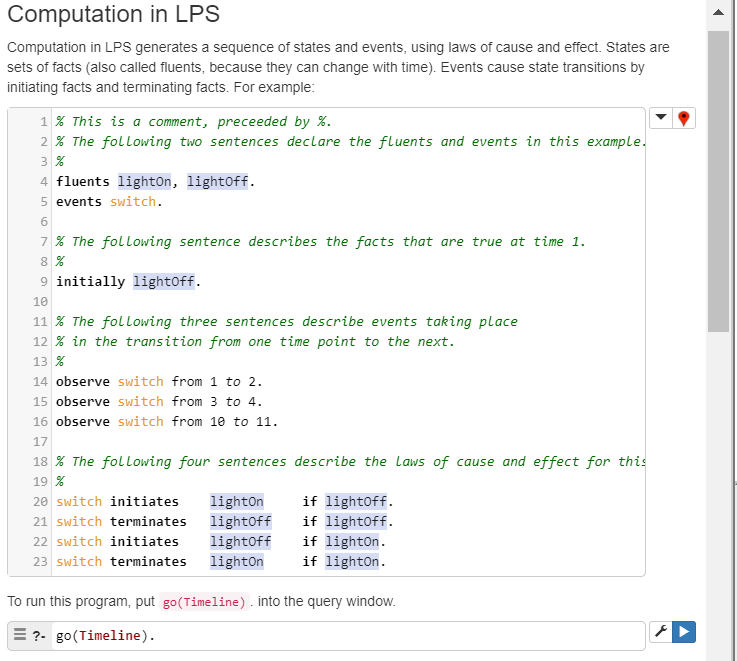}
	\caption{The first notebook in the Firststeps notebook in the Examples menu. The SWISH editor was extended, higlighting LPS-specific keywords, fluents and events.}
	\label{fig:lpsnotebook}
\end{figure}


The LPS interpreter previously ran on XSB and (SWISH-less) SWI Prolog. The SWISH implementation of LPS was developed to provide an easy-to-use online tool for teaching logic, computing and computational thinking. SWISH was chosen because it was the only Prolog system that provided the web-based interface and program editor needed for the task. Thanks to the SWISH and SWI-Prolog infrastructures, it was possible to develop the implementation with a limited budget of about four person-months. The open-source prototype is accessible from Imperial College London.\footnote{http://lps.doc.ic.ac.uk}

The multi-pane IDE interface and the notebook facility were not part of the original requirement, but were additional bonuses. The notebook facility of SWISH encouraged us to create an interactive tutorial, combining explanatory text with runnable and editable examples and queries. This is illustrated in \figref{lpsnotebook} by the first notebook in the First Steps with LPS notebook.

\begin{figure}
	\includegraphics[width=0.85\linewidth]{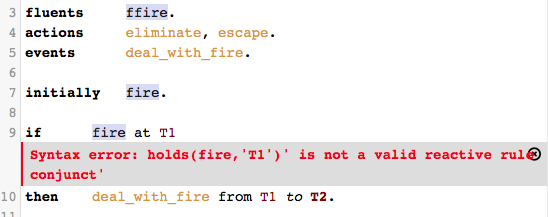}
	\caption{SWISH editor showing a LPS-specific error.}
	\label{fig:lpserror}
\end{figure}

Lines 4 and 5 of the program in the notebook are declarations. Line 4 declares that lightOn and lightOff are fluents, and line 5 declares that switch is an event. To aid in understanding the program, fluents are highlighted blue/grey in the SWISH program pane, and events are highlighted orange.

Line 9 states that the fluent lightOff is true in the initial state, at time 1.  Lines 14-16 record observations of external events, which occur in the transition from one state to the next. These events are “timestamped” by the “times” of the previous state and the next state. Lines 20-23 are causal laws, which specify the fluents that are initiated by an event and added to the current state, and the fluents that are terminated and deleted from the state. Events can also include actions generated by the program.

The fluent and event declarations help to detect syntactic errors, which are highlighted in red. The error message in \figref{lpserror} reports that fire at ${T1}$ (internally represented as ${holds(fire, T1)}$) is not a valid conjunct for the reactive rule in lines 9 and 10, because fire has not been declared as a fluent.

Given an initial state, and stream of external events and actions, computation in LPS updates the current state, using the causal laws. This is illustrated by the execution of the program in \figref{lpsnotebook}, which visualises the model as the timeline in \figref{lpstimeline}. The timeline is a graphical rendering of the term generated as an answer to the Prolog query ${go(Timeline)}$. Alternatively, models can also be presented in textual form as a list of states and events, using the nullary query $go$.

Timelines display models in their totality, from the initial state to the final state, including all states and events in between. In theory, time can be infinite. But in the implementation, the final state, which can be overridden, is at time 20 by default.

The program in \figref{lpsnotebook} illustrates a passive agent without any goals and without the ability to perform any actions. The second notebook in the same First Steps notebook illustrates an active agent, which can perform the action of switching the light (on or off) for itself, and which has the goal of switching the light as soon as it observes that the light is off:

\begin{figure}
	\includegraphics[width=\linewidth]{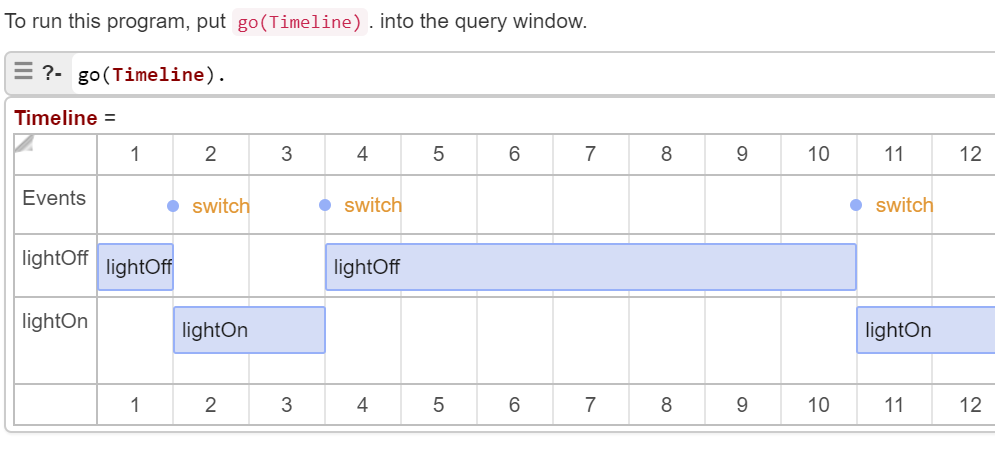}
	\caption{LPS execution timeline.}
	\label{fig:lpstimeline}
\end{figure}

\begin{code}
	if lightOff at T1 then switch from T1 to T2.
\end{code}
Here $T1$ and $T2$ are ``time'' (or state) variables. $T1$ is implicitly universally quantified and $T2$ is implicitly existentially quantified. The quantifiers are not written explicitly, because all variables in the antecedent of a rule are universally quantified, and all variables that are only in the consequent are existentially quantified. Because switch is an atomic action, the only instantiation of the variable \textit{T2} that solves the goal \textit{switch from T1 to T2} is where \textit{T2} is the next time after \textit{T1}.

\begin{figure}
	\includegraphics[width=\linewidth]{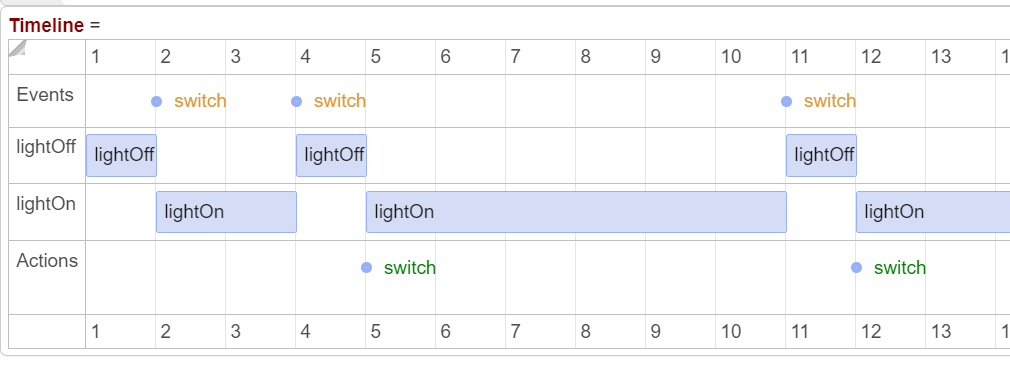}
	\caption{LPS execution timeline for the same program and same external events, but with a reactive rule if $lightO\!f\!f$ at $T1$ then $switch$ from $T1$ to $T2$.}
	\label{fig:lpstimeline2}
\end{figure}

The explicit representation of time clarifies the meaning of sentences. However, as a shortcut, time can often be left implicit. For example, the rule in this example can also be written more simply in the form:

\begin{code}
	if lightOff then switch.
\end{code}
No matter how the rule is written, for the rule to be effective, switch needs to be declared as an action. As a result, when the light is observed to be off at 4, the rule generates an action of switching the light between 4 and 5. See \figref{lpstimeline2}.

The first two notebooks illustrate agents whose only ``beliefs'' are ``facts'' or causal laws. However, beliefs in LPS can also include logic programs that define abstract concepts in terms of more concrete concepts, and logic programs that define complex events and plans in terms of other events, actions and fluents. The third notebook in First Steps illustrates a logic program defining an intensional predicate $happy$ in terms of extensional predicates $lightOn$ and $lightO\!f\!f$:

\begin{code}
	happy(bob) if lightOn.
	happy(dad) if lightOff.
\end{code}
In LPS, only extensional predicates are updated explicitly. Intensional predicates become true or false indirectly as ramifications of updates of extensional predicates. Times in the definition of an intensional predicate do not need to be written explicitly, because they are all identical.

In theory, the logical semantics of LPS allows any sentence of first-order logic to play the role of a goal. In this respect, the resulting combination of logic programs and first-order logic is like their use for model expansion in FO(ID) \cite{denecker2000}. The main difference is the focus in LPS on generating models that are sequences of states and events, using destructive updates as in imperative computer languages.

In contrast with the logical semantics of LPS, which allows arbitrary goals in first-order logic, the SWISH implementation allows only reactive rules and constraints. Constraints are written in the form \emph{false conditions}, where $conditions$ is a conjunction of events that all occur at the same time and fluents that hold before or after the events. In the fourth notebook, this is illustrated by the constraint:

\begin{code}
	false goto(dad, Place1), goto(dad, Place2), Place1 \= Place2.
\end{code}
which prevents dad from going to more than one place at a time. The constraint is necessary, because dad's  behaviour is governed by reactive rules that make him go to a room whenever there is a light on in the room, and to turn the light off if it is still on. In this example, times do not need to be written explicitly, because of the restrictions on times in constraints.

\begin{figure}
	\includegraphics[width=\linewidth]{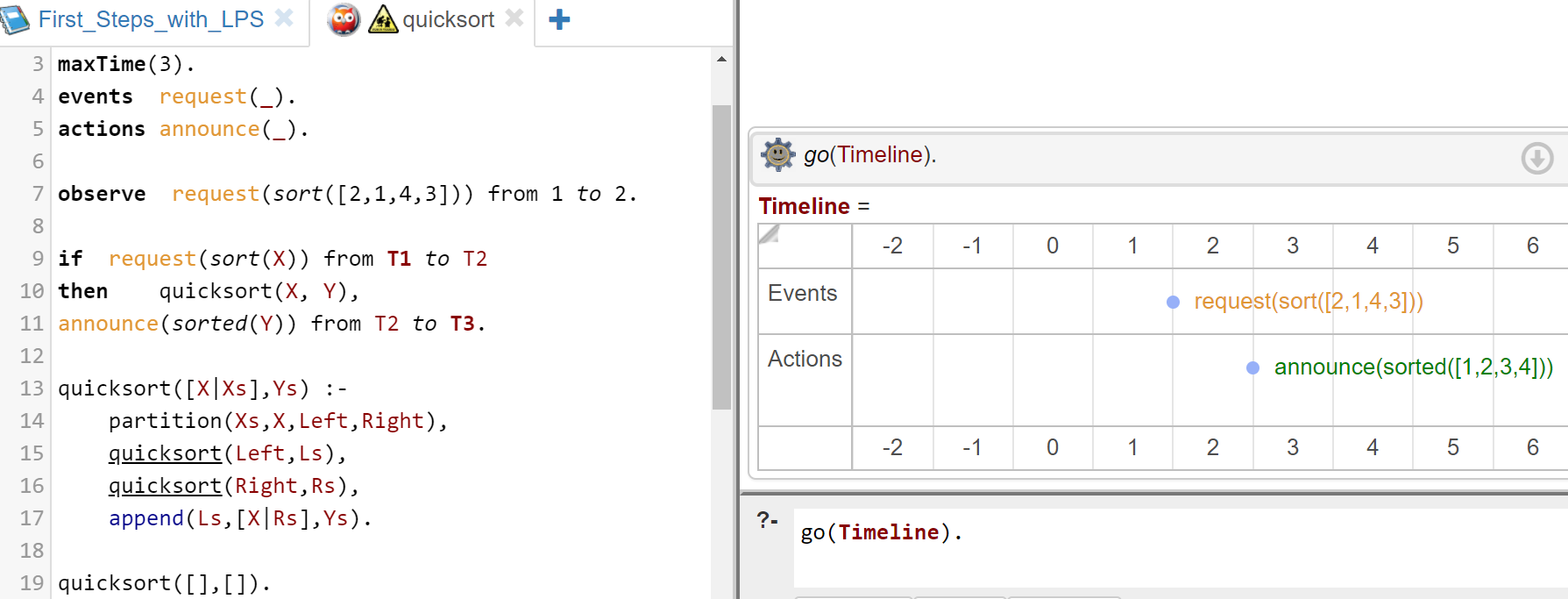}
	\caption{Timeless, auxiliary predicates can be defined using Prolog.}
	\label{fig:quicksort}
\end{figure}

Logic programs in LPS are pure Prolog programs, but are represented by Prolog facts, and are executed by a Prolog meta-interpreter. The meta-interpreter implements destructive updates, with its logical model generation semantics, using Prolog's assert and retract, which do not have a logical semantics. The use of Prolog to implement LPS means that auxiliary predicates in LPS can be defined by a compiled Prolog program, as in the example in \figref{quicksort}. The example also shows how LPS gives a simple, logical semantics to input-output, representing inputs as observations of external events, outputs as actions, and the relationship between inputs and outputs as a reactive rule. If the input is a request, then the rule may optionally include additional arguments and conditions to determine, for example, whether to fulfil the sender's request, reply with a polite refusal, or block future requests from the sender.

Models in LPS have two forms: A static form, which is needed for the logical semantics, and in which fluents and events are included in one, all-encompassing structure with timestamps; and a dynamic form, which is needed for practical implementation, and in which only the current state and events are stored without timestamps. Timelines display models in their static form. But models can also be displayed in their dynamic form, as animations.

\begin{figure}
	\includegraphics[]{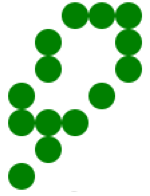}
	\caption{LPS 2D animation frame for the Game of Life.}
	\label{fig:life}
\end{figure}

For this purpose, we have used the rendering facilities of SWISH to develop a preliminary, declarative language for users to generate their own 2D animations. \figref{life} illustrates an animation frame representing a single, current state in Conway's game of life, in the LPS examples notebook in the examples menu.

\begin{figure}
	\includegraphics[width=0.60\linewidth]{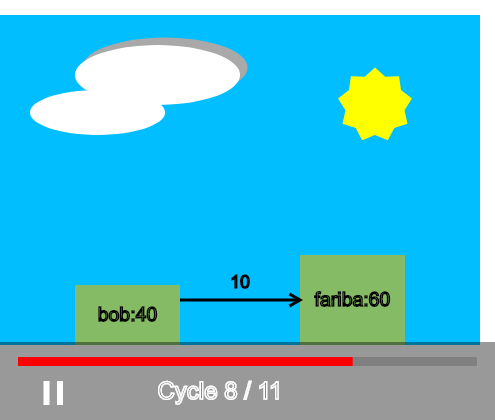}
	\caption{Animation of the bank account example.}
	\label{fig:bank2}
\end{figure}

To obtain the animation, it suffices to specify the visual representation of fluents and their 2D co-ordinates. For example, the Prolog clause:

\begin{code}
	d(X-Y,[center:[XX,YY], radius:5, type:circle, fillColor:green]) :-
	XX is X*10, YY is Y*10.
\end{code}

\noindent expresses that the fluent \emph{X-Y}, which represents the presence of a live cell at location X-Y, is displayed as a green circle centred at $[X*10, Y*10]$ with radius 5. The transition from one state to the next is specified by reactive rules, such as the rule that a live cell dies if it has less than two live neighbors:

\begin{code}
	if X-Y at T, aliveNeighbors(X-Y,N) at T, N<2 then die(X-Y) from T.
\end{code}
2D animation can be used for a wide range of examples. \figref{bank2} illustrates an animation frame for a bank transfer simulation in the same LPS example notebook. The video-like controls allow the user to stop and restart the animation at any time. In fact, as the figure suggests, time in LPS is more like a state during which time stands still in an observe-think-act agent cycle.

The changing state of bob's and fariba's bank accounts is controlled by reactive rules, designed to enrich bob at fariba's expense:

\begin{code}
	if   transfer(fariba, bob, X) from T1 to T2, 
	     balance(bob, A) at T2,   A >= 10
	then transfer(bob, fariba, 10) from T2 to T3.
	
	if   transfer(bob, fariba, X) from T1 to T2, 
	     balance(fariba, A) at T2,  A >= 20
	then transfer(fariba, bob, 20) from  T2 to T3.
\end{code}

The SWISH implementation of LPS is currently being used in an introductory AI course at Imperial College, and to provide self-study tutorials for more advanced students, some of whom have gone on to reimplement LPS in Python and JavaScript, and to use LPS in such applications as games.   

The online prototype has also attracted a wide range of external users. The bank transfer example, in particular, has attracted users with interests in financial technology and smart contracts. Partly in response to this interest and building on previous work \cite{bna} on representing legal documents as logic programs, the LPS team has created a start-up company, Logical Contracts\footnote{http://logicalcontracts.com} 
to represent legal documents in logical, executable form. LPS is well suited for such legal applications, because there is a natural correspondence between logic programs and constitutive rules, which define legal concepts, between reactive rules and regulative rules, which enforce legal obligations, and between constraints and prohibitions.

\section{Architecture}
\label{sec:arch}

SWISH consists of two parts. The \emph{client side}, running in a
browser, is implemented as a series of jQuery plugins, using Bootstrap\footnote{\url{https://getbootstrap.com/}}
for styling and RequireJS\footnote{\url{http://requirejs.org/}} for
package management. The \emph{server side} is completely implemented in
SWI-Prolog \cite{DBLP:journals/tplp/WielemakerHM08}. It builds on top of
the SWI-Prolog HTTP server libraries
\cite{DBLP:journals/tplp/WielemakerHM08}, the Pengines library
\cite{DBLP:journals/tplp/LagerW14} and the IDE support libraries that
provide data for auto completion, documentation and highlighting. In the
remainder of this section we first give a brief overview of the
available components, next we elaborate on some of these components and
finally we discuss the portability of the design.

\subsection{The client components}
\label{sec:client-components}

The client is written as a series of JavaScript jQuery plugins.
RequireJS is used to resolve the dependencies and, for production
installations, to compile all jQuery components into a single minified
file. Currently, \file{swish-min.js} is 1.1Mb (308Kb compressed) and
\file{swish-min.css} 188Kb (53Kb compressed).

\begin{itemize}
    \item The \textbf{program editor} is used to edit programs, queries,
    HTML and Markdown.  It is  described in
    in \secref{codemirror}.

    \item The \textbf{query editor} is derived from the code editor,
    providing additional access to examples, history and modifying the
    query (count solutions, time, order, distinct, etc.)

    \item A \textbf{runner} instance takes care of actually running
    a query.  It is created from zero or more source code fragments
    and a query.  It allows for controlling the query (ask for more
    solutions, abort or discard it) and collects the
    \jargon{answers}.  It can be asked to organise the answers in
    the traditional Prolog way as a series of \mbox{\arg{Var} =
    \arg{Value}} statements or as a \jargon{table}, providing a
    solution as a row of values.

    \item An \textbf{answer} instance is responsible for rendering
    a term that is part of the answer.  The baseline is a nested HTML
    \elem{span} reflecting the syntactic components of the term that
    allows for folding and unfolding terms. SWISH allows for alternative
    term rendering using, e.g., tables, charts, graphs, etc.  This may
    be compared to the traditional \index{portray/1}\predref{portray}{1} hook, but SWISH rendering can exploit
    the full capabilities of HTML5, CSS, SVG and JavaScript.  See
    \secref{rendering}.

    \item A \textbf{notebook}  instance is a list of \jargon{cells} built from
    the above components, see \secref{notebook}.  The notebook itself allows for adding,
    deleting and organising cells and provides the necessary code
    to serialise the content as an HTML document on the server and import other
    notebooks.

    \item  The social infrastructure makes other users aware that they
    are operating on the same program using an \jargon{avatar}.  The
    avatar has a notification area that indicates when a document is
    opened, closed or saved. Each document has an associated
    \textbf{chat room} that allows users to discuss code and share
    queries.
\end{itemize}

\subsection{The SWISH server components}
\label{sec:server-components}

The SWISH server is a monolithic web server written entirely in
SWI-Prolog. The code heavily depends on the SWI-Prolog web libraries
(see \secref{portability}). The current 20K source lines are distributed
over 67 files. Below we describe the key components:

\begin{itemize}
    \item The client side code editor is supported by a \textbf{highlight server}
    that produces semantically enriched highlighting information based
    on cross-reference analysis. In addition, the \jargon{mode
    declarations} from the documentation, e.g.,
    ``\exam{thread_create(:Goal, -Id, +Options) is det}'' are compiled into
    a JSON object to support CodeMirror
    insertion of \jargon{templates}. See \secref{codemirror}.

    \item Running a query is managed by a \textbf{Pengine} or  Prolog engine.  The
    Pengine library is part of SWI-Prolog
    \cite{DBLP:journals/tplp/LagerW14}.  \jargon{Sandboxing} queries
    is one of the optional facilities of the Pengine service and is
    discussed in \secref{sandbox}.

    \item Query answers are processed by the \textbf{rendering service}
    that is described in \secref{rendering}. A rendering plugin
    recognises a Prolog term (like \index{portray/1}\predref{portray}{1}) and emits an HTML node.
    Adding rendering plugins is a common way to customise
    SWISH. See \secref{extending}.

    \item Notebook \textbf{markdown} cells are supported by a server-side
    markdown engine that is borrowed from PlDoc, the SWI-Prolog
    documentation system \cite{Wielemaker:2007c}.

    \item The \textbf{storage} component stores documents in a format
    that is largely compatible with the GIT SCM.  The main difference
    is that documents are treated individually and not organized in a
    \jargon{tree}. The storage component is discussed in \secref{gitty}.

    \item  \textbf{Authentication} and \textbf{authorisation} can
    be used to provide optional or obligatory login using HTTP
    authentication (basic or digest) or oauth2 federated login.
    Authorised users can be assigned rights to documents.  Authorization
    can also be used to disable the \textbf{sandbox} restrictions on
    permitted queries for certain users.

    \item The \textbf{data source} plugin based service
    can be used to import data into SWISH, typically from the web.
    The default plugins allow loading CSV files, scraping web pages
    and running SPARQL queries.  In addition, Pengines can be used
    to access remote Pengine (or SWISH) servers.
\end{itemize}

\subsection{The program editor}
\label{sec:codemirror}

A proper editor is the most important component of a usable programming
environment. The editor must support the language, including syntax
highlighting, auto indentation, code completion based on templates and
already existing code, highlighting of errors and warnings from the
compiler and providing access to the documentation.

Prolog is a difficult language to support in code editors due to the
lack of reserved keywords, the fact that there is no distinction between
code and data and the ability to extend the syntax using new operators.
For example, the word \const{if} in C starts an if-statement if not
embedded in a comment or a string, but the word \const{is} in Prolog can
refer to the built-in predicate \index{is/2}\predref{is}{2}, but also to some predicate with a
different arity, to just a constant, etc. Another example is \exam{X-Y}
which can both be an arithmetic expression or a \jargon{pair} as used
within, e.g., \index{keysort/2}\predref{keysort}{2}. SWI-Prolog's native editor, called PceEmacs,
solves this problem by closely integrating Prolog with the editor.
While typing, the current term (clause or directive) is parsed and
analysed in the context of the current file and its imports after
each keystroke. If the term has valid syntax, all tokens are coloured
according to their syntactic category as well as their semantic role in the
program. For example, a call to a non-existing predicate is coloured
red, a call to a built-in or imported predicate is blue and a recursive
call is underlined. The libraries that implement this analysis have been
decoupled from the native IDE, both to support source colouring for the
SWI-Prolog documentation system PlDoc \cite{Wielemaker:2007c} and
ProDT.\footnote{\url{http://prodevtools.sourceforge.net}, these libraries
are not yet used by ProDT.}

There are two dominant open source and actively maintained in-browser
code editors available: ACE and CodeMirror. We opted for CodeMirror
because its highlighting is based on raw JavaScript code rather than a
regular expression-based template language as used by ACE. The low
level implementation allows for a novel highlighting implementation. The
highlighter consists of a Prolog \emph{tokeniser} in JavaScript.
Tokenising Prolog is sufficient to colour comments, quoted material
(strings, quoted atoms), variables and constants (atoms and numbers).
The tokeniser also keeps track of bracket nesting to support smart
indentation. CodeMirror's token-based highlighting does \emph{not} allow
for  lookahead over line breaks. As a result we are not able to count
the \jargon{arity} of terms that are broken over multiple lines. As
terms with the same name but different arity are not related in Prolog
this prevents unambiguous meaningful colouring.

We resolve this issue using server-side support to realise
\jargon{semantic} highlighting. First, the changes to the content of the
editor are forwarded to the server which maintains a mirror of the
editor content. Second, the editor asks the server to produce a list of
semantically enriched tokens for the source. The tokens are returned as
a list-of-lists, where each inner list represents the tokens for a
source term (clause or directive). Grouping the tokens per source term
allows for incremental update (not implemented) as well as
re-synchronisation (see below). For example, consider the following
program.

\begin{code}
go :-
    non_existing(X).
\end{code}

The JavaScript client produces the tokens [ \textit{atom},
\textit{atom}, \textit{functor}, \textit{punct}, \textit{var},
\textit{punct}, \textit{fullstop} ]. The server analyses the program
text and produces [ \textit{head (not called)}, \textit{neck},
\textit{goal (undefined)}, \textit{punct}, \textit{var (singleton)},
\textit{punct}, \textit{fullstop} ]. The JavaScript tokeniser examines
both its own tokens and the server token list. If the tokens are
compatible (for example `atom' is consistent with `head'), it uses the
enriched server information (head of a not-called predicate) to
determine the colour and font. If the tokens are not compatible (for
example the client finds \textit{var} and the server reports
\textit{atom}, it schedules a request to the server for a revised list
of enriched tokens. This request is sent if the user pauses typing for
two seconds. While waiting for up-to-date enriched tokens, the
JavaScript highlighting code heuristically tries to re-synchronise and
either uses the uncertain results or falls back to the plain tokens.
Heuristic re-synchronisation checks whether a single added, deleted or
modified token gets the token stream in-sync. If this fails it tries to
re-synchronise on a full-stop with the next clause or directive.

A CodeMirror \jargon{hover} plugin is used to show basic information
about tokens if the pointer hovers over it. For goals, this includes
where the goal is defined (ISO, SWI-Prolog built-in, a library, locally)
and the documentation summary information if available. This information
is requested from the server.

A CodeMirror \jargon{template} plugin is configured from templates
(e.g., \exam{atom_length(+Atom, -Length)}) extracted from the SWI-Prolog
manual and PlDoc documentation of imported libraries. This plugin shows
a menu of applicable predicates with their templates when the user
presses \textsf{Control-Space}.

Finally, if the compiler generates errors or warnings, these are
inserted as notes into the source code.

\subsection{Document storage and version management}
\label{sec:gitty}

SWISH explicitly targets the cooperative development of Prolog programs
and queries, optionally organised in notebooks. Cooperative development
requires version management. The storage service is implemented in
Prolog and inspired by the GIT SCM. Unlike GIT, each file is versioned
independently rather than maintaining the version of a hierarchy of
files because files are generally independent. The object representation
of our storage system is binary compatible with GIT. The content of a
\jargon{commit} object is a SWI-Prolog \jargon{dict} object linking the
current version to the previous one
and providing metadata such as the time,
author and commit message. Saving a new version adds a record to a
journal file that provides the file name, its previous commit hash and
the new commit hash. The file-based repository can be shared between
multiple SWISH processes that may run on different nodes if a shared
file system is used and the file system supports advisory locks for
synchronising writes to the journal file. The journal file facilitates
fast discovery of the available heads and fast tracking of changes by
other SWISH instances.  The ability to run multiple SWISH servers on
a shared or distributed file system improves its scalability.

Files can be referenced by content using their hash or by name. The name
can be considered a version \jargon{head} and refers to the latest
version with that name. The file interface provides the following
operations:

\begin{itemize}
    \item Saving a file anonymously, which produces a randomly
          generated URL.
    \item Saving a file by name.
    \item Organise files using \jargon{tags}.
    \item Saving a new version.
    \item Merge changes if another user saved the same file.
    \item Show the available versions and modifications between
          versions.
    \item Forking a file under a new name.  The history remains
          linked to the original.  Both the old and new name act
	  as a revision \jargon{head}.
\end{itemize}

Prolog source files can include other programs on the same server using
\exam{:-~include(filename).}, including the latest version or
\exam{:-~include(hash).} to include a specific version.

\subsection{Server side execution of the query}
\label{sec:pengines}

Server-side execution of a query is supported by the Pengines
\cite{DBLP:journals/tplp/LagerW14} library. The Pengines library allows
for creating a Prolog engine possibly on a remote server represented by
a Prolog thread. Optionally, the Pengine is handed a Prolog program that
is loaded into the Pengine's workspace, a Prolog module. The Pengine may
be asked queries through HTTP, similar to a traditional Prolog user
interacting with Prolog running in a
terminal.\footnote{\url{https://www.youtube.com/watch?v=G_eYTctGZw8}}

\subsection{Rendering answers using HTML}
\label{sec:rendering}

The \jargon{rendering} infrastructure exploits one of the key benefits
of using web technology, the ability to use arbitrary HTML, CSS,
Javascript and SVG to visualise answers. \Figref{render} and
\figref{lpstimeline} provide  examples. A rendering \jargon{plugin}
is a Prolog module that exports a Prolog grammar rule (DCG) named
\index{term_rendering//3}\dcgref{term_rendering}{3}. This rule
is passed the term to be rendered, a list of \arg{Name}~=~\arg{Var}
bindings providing the names of variables as provided by
\index{read_term/2}\predref{read_term}{2} and a list of options. If the
renderer recognises a Prolog term, it produces a list of HTML tokens
following the rules of SWI-Prolog DCG-based HTML generation
\cite{DBLP:journals/tplp/WielemakerHM08}. The rendering plugin is
registered using a directive
\index{register_renderer/2}\predref{register_renderer}{2}, which defines
the name of the renderer and a short description. A rendering plugin is
activated for a program using the directive below. The \arg{Options}
term is optional. If provided it is passed to
\index{term_rendering//3}\dcgref{term_rendering}{3} in its third
argument.

\begin{code}
:- use_rendering(Renderer, Options).
\end{code}

\noindent
Each term is handed to all active renderers and finally to the
built-in generic term renderer that represents any term as \index{writeq/1}\predref{writeq}{1}
using nested \elem{span} elements. If one or more of the special purpose
renderers succeed, the output of the first is displayed, together with a
hover-menu that provides access to the other representations.

Currently, rendering plugins are not executed inside the Pengine, but by
the HTTP handler that returns a Pengine answer to the client. This is a
consequence of the Pengine interface that will be addressed in future
versions as it prevents the renderer from accessing global data,
bypasses resource limitations imposed on the Pengine and prevents
supporting rendering hooks as part of the user program.

\subsection{Sandboxing queries}
\label{sec:sandbox}

Prolog systems provide a programming environment that is capable of
changing itself permanently and which exposes a potentially dangerous
interface to the operating system. Both for education purposes and data
analysis though, one can write meaningful programs without making
permanent changes to the server or the server's file system. Preventing
the user from performing dangerous operations is the task of the sandbox
library. The sandbox is active while loading a program into the
Pengine's program space, where it refuses to add clauses to other
modules and where it only accepts a restricted set of
\jargon{directives}, also aimed at keeping all changes local to the
workspace. Prior to execution, the sandbox unfolds the query and
compares all reachable goals with a whitelist. It fails under one of
these conditions:

\begin{itemize}
    \item It discovers a (meta-) goal for which it cannot deduce the
    called code.  The traditional example is \exam{read(X), call(X)}.
    If such a goal is encountered, it signals an \jargon{instantiation
    error}, together with a trace that explains how the insufficiently
    instantiated goal can be reached. Note that it can deal with normal
    high-order predicates if the meta-argument is specified. For
    example, the following goal is accepted as safe.

    \begin{code}
    ?- maplist(plus(1), [1,2,3], List).
    \end{code}

\noindent
\noindent
    \item It discovers a goal that is not whitelisted.  In this case
    it signals a \jargon{permission error}, again accompanied with
    a trace that explains how the goal can be reached.  Note that
    pure Prolog predicates are unfolded even  if they are
    built-in or library predicates.
    Unfolding stops if a predicate is in the whitelist.

    \item It discovers a cross-module  call (\arg{M:Goal})  to a predicate
    that is not public. Normally, SWI-Prolog, in the tradition of
    Quintus Prolog, allows for this. Allowing it in SWISH would imply
    that no data can be kept secret. With this  prohibition,
    libraries can
    keep data in local dynamic predicates that remain invisible to
    non-authorised users.
\end{itemize}

\subsection{Remote access to SWISH}
\label{sec:remote}

SWISH can be integrated in workflows with external tools and data. The
integration options are embedding, downloading query results as CSV
(comma separated values),
accessing SWISH as a service and accessing other services from SWISH.

SWISH  can be \jargon{embedded}  in a web page using an \elem{iframe}.
This is
particularly appealing for educational deployment: textual material
can be interleaved with SWISH frames. The SWISH interface
can be preloaded with a \jargon{background program} (an invisible
program that is sent along with each query), a \jargon{program} (that
appears in the code editor), a \jargon{query} and \jargon{examples} that
appear in the query editor. It can also be preloaded with a
\jargon{notebook}.  See \secref{lpn}.

Both a \jargon{runner} and the notebook query editor provide a menu to
download the query answers as CSV. The subsequent dialogue allows for
setting a limit on the number of results and defining the
\jargon{projection}, the variables to be  included in the CSV and the order of
these variables.

SWISH offers the Pengine API (Application Programming Interface) to clients in Prolog, JavaScript, Java, Ruby and
Erlang. This API can be used to execute queries against programs that
are provided by the client or available as saved documents on the SWISH
server.

The client for Prolog offers a predicate for making non-deterministic
RPC (remote procedure calls), for example:

\begin{code}
?- use_module(library(pengines)).
true.
?- pengine_rpc('https://swish.swi-prolog.org', member(X,[a,b,c])).
X = a ;
X = b ;
X = c.
?-
\end{code}

\subsection{SWISH Implementation of cplint}
\label{sec:cplintImpl}
All the \cplintonswish{} algorithms are available as modules to be loaded in a regular Prolog
input file. Each module offers directives to separate various sections of the
source with different roles: for example, for holding the probabilistic logic
program for inference or parameter learning or the background knowledge for
parameter or structure learning.
These directives are handled using \index{term_expansion/2}\predref{term_expansion}{2} and activate or deactivate
other \index{term_expansion/2}\predref{term_expansion}{2} clauses that process the probabilistic clauses and
transform them into Prolog clauses to be used by the  algorithms.

The modules were adapted for the multi-user context
of \cplintonswish{}. The modules use asserts to store information in the
database: these asserts
are now performed in the module that is created for the individual
Prolog engine (see \secref{arch}).
 During the execution of
the algorithms all asserts are performed in the temporary Prolog engine
module and discarded at the end of the computation.

\subsection{SWISH Implementation of LPS}
\label{sec:lpsImpl}

The LPS program is preprocessed into Prolog using the system predicate \index{term_expansion/2}\predref{term_expansion}{2}. In addition, the interpreter was made thread-safe (wrt to SWISH's multithreaded server environment) by declaring relevant dynamic predicates with SWI Prolog's \index{thread_local/1}\predref{thread_local}{1} directive.

To help with reading and understanding an LPS program, the LPS specific syntax (of fluents, events and actions, as well as keywords such as $if$, $then$ and $false$) is highlighted by plugging into and extending SWISH's syntax colouring framework. The LPS pre-processor also checks for semantic errors and flags them when they occur, as in \figref{lpserror}. For this purpose, the pre-processor keeps a record of LPS source positions, and, when an error occurs, it delegates a report of the error to SWI Prolog's \index{print_message/2}\predref{print_message}{2} predicate, passing an error term with the position. The SWISH machinery displays such messages as notes in the source code\footnote{Such notes are displayed by CodeMirror, but they are \emph{not} part of the user document.}

Each execution of a query $go(Timeline)$ generates a post-mortem trace as a set of Prolog facts, which is converted into a JSON representation, visualised by a JavaScript module.\footnote{http://visjs.org/docs/timeline} The timeline is implemented as a SWISH custom renderer.
Similarly, 2D animations are displayed via a SWISH custom renderer, developed using a 2D object library.\footnote{http://paperjs.org}

More recently, the start-up Logical Contracts has further enhanced the SWISH implementation to allow the execution of LPS programs in the background, providing  web services to client applications, which can query the background server for access to the server's fluents, receive events injected by the server, and delegate actions to other systems as REST requests. These events and actions can include interaction with the Ethereum blockchain.

\section{Related work}
\label{sec:related}

We are not aware of other initiatives that aim at developing a rich
web-based development environment for Prolog. There are web sites that
allow  running Prolog online such as Tutorial
Points.\footnote{\url{https://www.tutorialspoint.com/execute_prolog_online.php}}
These sites either run your Prolog program as a batch job or provide a
classical terminal to access Prolog. None of the environments we visited
provide SWISH features such as  notebooks or rendering Prolog
answers using web technology such as charts.

We do not compare SWISH with traditional editor or GUI based development
environments for Prolog such as GNU-Emacs or
ProDT,\footnote{\url{http://prodevtools.sourceforge.net/}} because
web-based environments provide both new opportunities such as shared
access and pose new challenges such as a more limited interface, limited
bandwidth and high latency. Instead, we discuss three applications
that  served as inspiration for SWISH:
JSFiddle,\footnote{\url{https://jsfiddle.net/}}
R-Studio\footnote{\url{http://www.rstudio.com/}} and Jupyter
Notebooks.\footnote{\url{http://jupyter.org}}

\begin{itemize}
\item
The initial inspiration for SWISH was JSFiddle. JSFiddle is
an environment for  testing browser technology (HTML, CSS, JavaScript)
and thus it is naturally executed inside the user's browser. SWISH programs
are executed on the server. For educational purposes, client side
execution is probably feasible. For data analysis purposes remote
execution allows users to control large Prolog jobs on big servers from
their browser.

\item
R-Studio \cite{gandrud2013reproducible} is an interface to the R
statistical package. Although not a web application, it is based on the
Qt webkit framework and uses web based technology in the background. The
R-studio interface has a similar layout as SWISH, providing a source
window, a console and an output window that typically shows results as
tables or charts.

\item
Jupyter notebooks (formerly IPython notebooks)
\cite{rossant2013learning} allow mixing markdown or HTML text with
Python  code. When executed, the notebook shows the text, sources and
possible results as numbers, tables or charts.
\end{itemize}

SWISH embodies many of the ideas behind JSFiddle, R-Studio and Jupyter.
SWISH provides shared anonymous access like JSFiddle. Both the R-Studio
`program and output window' and Jupyter notebook interfaces are
available as they serve different user communities in our experience.

Both R-Studio and Jupyter notebooks work on the basis of
\jargon{authenticated access}. Once access is obtained, any command may
be executed. SWISH can operate both as a public service granting access
to non-intrusive queries that may be executed concurrently on shared
(typically read-only) data and as an authenticated service to run
arbitrary queries. The authenticated mode is often used for maintenance
purposes such as updating the server software or loading data for
further shared analysis.

Where Jupyter supports other languages using the notion of Jupyter
\jargon{kernels}, such support is not provided by SWISH. We believe that
Prolog is sufficiently different from the languages targeted by Jupyter
to justify dedicated support. SWISH has little to offer to compete with
Jupyter on the languages that Jupyter supports properly though.

Unlike JSFiddle,
R-Studio and Jupyter, SWISH can easily be embedded into web pages using an
\elem{iframe} element where the program, examples and queries can be
provided in the URL that loads SWISH. This feature can be used to
transform source code in a static HTML page into a SWISH instance using
a simple JavaScript call. This is demonstrated in \secref{lpn}.

\cplintonswish{} is a system for PLP that
uses SWISH to provide a web interface to \cplint{}.
A related PLP system is ProbLog2
\cite{DBLP:journals/corr/abs-1304-6810}, which has also an online
interface\footnote{\url{https://dtai.cs.kuleuven.be/problog/}} \cite{Dries2015}.  The
main difference between the web interfaces of \cplintonswish{} and ProbLog2
is that the first uses a Prolog-only software stack, whereas the latter
 relies
on several different technologies, including Python 3 and
external knowledge compilers.
ProbLog2 can also be used in Jupyter notebooks
but, differently from SWISH, cannot interface to R.

\section{Discussion}
\label{sec:evaluation}

We have not performed any direct user evaluations of SWISH. Given the
pace of development, different configurations and user communities, it is
not feasible to perform such a study with limited resources. Below we
summarise the impact of the system, including  some lessons learned regarding using
Prolog through a web app, and discuss
 portability.

SWISH  has been
online (September 2017) for three years. It
is extensively being used for Prolog education,
 as discussed in \secref{edu} and \secref{extending}.
Since its launch, at
least three independently developed educational sites have been launched:
\cplint{} and LPS (described in this article) and the book
\textit{Simply Logical} \cite{flach1994simply}. The main server now has
35,420 programs stored with 68,806 revisions. Usage is strongly
correlated with academic holidays, ranging from 36,415 queries per week
(July~31 - August~6, 2017) to 125,946 (May~7-14, 2017). Google link
search (\verb$link:swish.swi-prolog.org$) reports 8,490 links. The
popularity of the Github repository, 198 stars and 48 forks compared to
the main SWI-Prolog repository with 223 stars and 53 forks is another
indication of the impact of SWISH.

The development of SWISH and its usage as a public service 
demonstrates  SWI-Prolog's ability for writing web
services. Moreover, it represents an
excellent stress-test for SWI-Prolog. We have been forced to
extend resource management, e.g., by limiting the program space
associated with the temporary modules 
to avoid extensive use of \index{assert/1}\predref{assert}{1} causing
 the server to run out of memory. Timeout and
exception handling needed to be improved to avoid runaway queries.
In particular, resource exceptions such as stack
overflows are frequent and
need to be handled reliably. Both scalability and reliability of
multi-threaded execution needed to be improved. We have now reached a level
of stability were restarts due to software upgrades (every one to two
weeks) are more frequent than restarts due to crashes, memory leaks or runaway queries. We have
shown that  SWI-Prolog-based servers are capable of concurrently running arbitrary
queries with bounds on resources.

The current browser infrastructure has proven to be sufficiently rich,
portable and performant for developing a web IDE for Prolog.
Server-assisted semantic highlighting has proven to be feasible, but the
mechanism is rather fragile because it requires synchronisation of two
independent highlighting processes, one at the client and one on the
server. Besides providing shared access, a web front end allows for
transforming Prolog answers into rich graphical representations such as
tables, charts, graphs or the LPS timelines.

\subsection{Portability}
\label{sec:portability}

The SWISH client libraries are based on mature JavaScript libraries. The
client runs in all modern major browsers with HTML5, CSS and JavaScript
support. It is frequently tested on Firefox, Chrome, Safari and Internet
Explorer~11.

The server code is basically non-portable. Many of the required
libraries and features are shared with at least one other Prolog
implementation, but none is capable to support the full range. Below we
summarise the main issues.

\begin{itemize}
    \item The scale of the involved Prolog libraries demands 
    a closely compatible Prolog module system.  Probably only
    SICStus and YAP can be used without significant restucturing.

    \item The HTTP server libraries are heavily based on C~code
    that interacts with the SWI-Prolog foreign language interface
    to Prolog streams.  YAP has copied the low-level libraries and is
    capable to run an old version of these libraries.

    \item The Pengines library depends on the HTTP library and
    the multi-thread interface.  The SWI-Prolog thread API is
    also provided by YAP and XSB.

    \item The sandbox library (\secref{sandbox}) assumes that
    whitelisted predicates both have no side effects and are robust
    against stack overflows, cyclic terms, etc. Few Prolog
    systems can satisfy this requirement. SICStus Prolog would be a
    candidate.

    \item The semantic syntax highlighting depends on detailed
    source layout information provided by \index{read_term/3}\predref{read_term}{3}.  SWI-Prolog's
    support for term layout is an extended version of the Quintus
    Prolog term layout functionality.

    \item Significant parts of the code rely on SWI-Prolog version~7
    extensions, notably the \jargon{dict} and \jargon{string} types
    that facilitate a natural mapping between Prolog and JSON data.
\end{itemize}

From the above list it should be clear that a fully functional port of
SWISH to another Prolog system is not immediately feasible.  There is a
more realistic scenario though. In this setup, SWI-Prolog provides the
web interface and most of the development tools and another language,
not even necessarily Prolog, provides the query solving. The interface
between the two can be based on inter-process communication or, if the
target system is safe and capable of supporting threads, by linking the
target system into the process and using the C~interface.

Inter-process communication is already used in SWISH for embedding
R.\footnote{\url{http://www.swi-prolog.org/pack/list?p=rserve_client}}
This embedding is based on
Rserve\footnote{\url{https://cran.r-project.org/web/packages/Rserve/}}.
Rserve consists of an R instance that is fully initialised and
optionally contains pre-loaded libraries. The R server provides a TCP/IP
or Unix socket and a binary protocol to manage sessions and R commands.
A new connection causes the server to fork a new child that handles the
client. Resource management and isolation are provided by running Rserve
inside a Docker based OS sandbox.\footnote{\url{https://github.com/JanWielemaker/rserve-sandbox}}

\subsection{Scalability}
\label{sec:scalability}

The SWISH server is a monolithic SWI-Prolog program. Some parts of the
interaction with SWISH are by nature stateful, notably query execution
and syntax highlighting, which maintains a clone of the user's editor on
the server. These parts can be distributed over a pool of
servers using a proxy server with support for \jargon{sticky sessions}.
Sticky session support implies that the proxy assigns a new user to a
random server and subsequently forwards all trafic from this user to the
same server. Other parts, such as serving the main page, icons, CSS,
JavaScript and manual pages can easily be offloaded from the Prolog
server by using a caching reverse proxy or serving them from another
server. Secondary servers may also be used to offload the chat and
storage services. Extrapolating from resource usage on the main public
server we estimate that a single SWISH server can handle up to about
1,000 concurrent users (dual Intel Xeon E5-2650) for a typical
educational workload.

\section{Conclusions}
\label{sec:discussion}

This article presented SWISH, a web service that supports the execution of
Prolog and domain specific languages (DSLs) defined on top of Prolog
from a browser. SWISH provides an interface that is primarily targeted
at program development and testing as well as a \jargon{notebook}
interface that is argeted at capturing workflows for data
analysis and tutorials. The SWISH server is a SWI-Prolog program
where user queries are executed by Pengines which are
built from a Prolog thread and an isolated temporary Prolog module. The
monolithic approach provides easy deployment as well as shared access
to, e.g., large pre-loaded Prolog knowledge bases.

The current interface provides a better interaction with and more
control for managing queries as well as small Prolog programs than the
traditional Prolog command line and Prolog IDEs. This makes it suitable
for education and data analysis. It is not yet suitable for the
development of large applications. For these scenarios, users expect
support for modular program development, a richer editor and a better
debugger.

The current system does not provide much support for animating progress,
such as showing the moves while solving the towers of Hanoi puzzle. One
option used in LPS is to generate the entire animation and then use
JavaScript to play the animation in the user's browser. Future versions
may exploit the Pengine \jargon{prompt} mechanism that is also used to
implement \index{read/1}\predref{read}{1} to read from the browser,
 for creating interactive
animations. Currently, SWISH rendering plugins are executed in the
context of the HTTP reply mechanism rather than by the Pengine itself.
This implies that they are not sandboxed, they must be provided by the SWISH
installation and we cannot support user defined rendering plugins.

In data analysis scenarios some queries take long to execute and, if we
provide a result, we would like to be able to reliably reproduce this
result and apply exactly the same version of the analysis program to new
data. For this purpose we are developing a persistent answer cache based
on a hash of the called predicate and its dependencies.

\section*{Acknowledgements}

This research was partially supported by the VRE4EIC project, a project
that has received funding from the European Union's Horizon 2020
research and innovation program under grant agreement \textnumero
676247. The SWISH implementation of LPS was developed under an EPSRC
grant administered by Imperial College London. We thank the referees for their helpful comments.

\bibliography{swish}

\begin{thebibliography}{}

\bibitem[\protect\citeauthoryear{Alberti, Bellodi, Cota, Riguzzi, and
  Zese}{Alberti et~al\mbox{.}}{2017}]{AlbBelCot17-IA-IJ}
{\sc Alberti, M.}, {\sc Bellodi, E.}, {\sc Cota, G.}, {\sc Riguzzi, F.}, {\sc
  and} {\sc Zese, R.} 2017.
\newblock \texttt{cplint} on {SWISH}: Probabilistic logical inference with a
  web browser.
\newblock {\em Intelligenza Artificiale\/}~{\em 11,\/}~1, 47--64.

\bibitem[\protect\citeauthoryear{Bellodi and Riguzzi}{Bellodi and
  Riguzzi}{2013}]{BelRig13-IDA-IJ}
{\sc Bellodi, E.} {\sc and} {\sc Riguzzi, F.} 2013.
\newblock Expectation {Maximization} over binary decision diagrams for
  probabilistic logic programs.
\newblock {\em Intelligent Data Analysis\/}~{\em 17,\/}~2, 343--363.

\bibitem[\protect\citeauthoryear{Bellodi and Riguzzi}{Bellodi and
  Riguzzi}{2015}]{BelRig15-TPLP-IJ}
{\sc Bellodi, E.} {\sc and} {\sc Riguzzi, F.} 2015.
\newblock Structure learning of probabilistic logic programs by searching the
  clause space.
\newblock {\em Theory and Practice of Logic Programming\/}~{\em 15,\/}~2,
  169--212.

\bibitem[\protect\citeauthoryear{Blackburn, Bos, and Striegnitz}{Blackburn
  et~al\mbox{.}}{2006}]{blackburn2006learn}
{\sc Blackburn, P.}, {\sc Bos, J.}, {\sc and} {\sc Striegnitz, K.} 2006.
\newblock {\em Learn prolog now!} Vol.~7.
\newblock College Publications, London, UK.

\bibitem[\protect\citeauthoryear{Byrd}{Byrd}{1980}]{Byrar}
{\sc Byrd, L.} 1980.
\newblock {Understanding the Control Flow of Prolog Programs}.
\newblock In {\em Logic Programming Workshop}. Department of Artificial
  Intelligence, University of Edinburgh, Debrecen, Hungary.

\bibitem[\protect\citeauthoryear{De~Raedt and Kimmig}{De~Raedt and
  Kimmig}{2015}]{DBLP:journals/ml/RaedtK15}
{\sc De~Raedt, L.} {\sc and} {\sc Kimmig, A.} 2015.
\newblock Probabilistic (logic) programming concepts.
\newblock {\em Machine Learning\/}~{\em 100,\/}~1, 5--47.

\bibitem[\protect\citeauthoryear{{De Raedt}, Kimmig, and Toivonen}{{De Raedt}
  et~al\mbox{.}}{2007}]{DBLP:conf/ijcai/RaedtKT07}
{\sc {De Raedt}, L.}, {\sc Kimmig, A.}, {\sc and} {\sc Toivonen, H.} 2007.
\newblock {ProbLog}: A probabilistic {Prolog} and its application in link
  discovery.
\newblock In {\em 20th International Joint Conference on Artificial
  Intelligence, Hyderabad, India (IJCAI-07)}, {M.~M. Veloso}, Ed. Vol.~7. AAAI
  Press, Palo Alto, California USA, 2462--2467.

\bibitem[\protect\citeauthoryear{Denecker}{Denecker}{2000}]{denecker2000}
{\sc Denecker, M.} 2000.
\newblock Extending classical logic with inductive definitions.
\newblock In {\em Computational logic—CL 2000}. Springer, Cham, 703--717.

\bibitem[\protect\citeauthoryear{{Di Mauro}, Bellodi, and Riguzzi}{{Di Mauro}
  et~al\mbox{.}}{2015}]{DBLP:journals/ml/MauroBR15}
{\sc {Di Mauro}, N.}, {\sc Bellodi, E.}, {\sc and} {\sc Riguzzi, F.} 2015.
\newblock Bandit-based monte-carlo structure learning of probabilistic logic
  programs.
\newblock {\em Machine Learning\/}~{\em 100,\/}~1, 127--156.

\bibitem[\protect\citeauthoryear{Dries, Kimmig, Meert, Renkens, Van~den Broeck,
  Vlasselaer, and De~Raedt}{Dries et~al\mbox{.}}{2015}]{Dries2015}
{\sc Dries, A.}, {\sc Kimmig, A.}, {\sc Meert, W.}, {\sc Renkens, J.}, {\sc
  Van~den Broeck, G.}, {\sc Vlasselaer, J.}, {\sc and} {\sc De~Raedt, L.} 2015.
\newblock Problog2: Probabilistic logic programming.
\newblock In {\em Machine Learning and Knowledge Discovery in Databases:
  European Conference, ECML PKDD 2015, Porto, Portugal, September 7-11, 2015,
  Proceedings, Part III}, {A.~Bifet}, {M.~May}, {B.~Zadrozny}, {R.~Gavalda},
  {D.~Pedreschi}, {F.~Bonchi}, {J.~Cardoso}, {and} {M.~Spiliopoulou}, Eds.
  Springer International Publishing, Cham, 312--315.

\bibitem[\protect\citeauthoryear{Fierens, den Broeck, Renkens, Shterionov,
  Gutmann, Thon, Janssens, and {De Raedt}}{Fierens
  et~al\mbox{.}}{2015}]{DBLP:journals/corr/abs-1304-6810}
{\sc Fierens, D.}, {\sc den Broeck, G.~V.}, {\sc Renkens, J.}, {\sc Shterionov,
  D.~S.}, {\sc Gutmann, B.}, {\sc Thon, I.}, {\sc Janssens, G.}, {\sc and} {\sc
  {De Raedt}, L.} 2015.
\newblock Inference and learning in probabilistic logic programs using weighted
  boolean formulas.
\newblock {\em Theory and Practice of Logic Programming\/}~{\em 15,\/}~3,
  358--401.

\bibitem[\protect\citeauthoryear{Flach}{Flach}{1994}]{flach1994simply}
{\sc Flach, P.} 1994.
\newblock {\em Simply Logical: Intelligent Reasoning by Example}.
\newblock Wiley, Chichester, UK.

\bibitem[\protect\citeauthoryear{Gandrud}{Gandrud}{2013}]{gandrud2013reproducible}
{\sc Gandrud, C.} 2013.
\newblock {\em {Reproducible Research with R and R Studio}}.
\newblock CRC Press, Boca Raton, FL.

\bibitem[\protect\citeauthoryear{Knuth}{Knuth}{1984}]{knuth}
{\sc Knuth, D.~E.} 1984.
\newblock Literate programming.
\newblock {\em The Computer Journal\/}~{\em 27,\/}~2, 97--111.

\bibitem[\protect\citeauthoryear{Kowalski and Sadri}{Kowalski and
  Sadri}{2015}]{lpsmodelgeneration}
{\sc Kowalski, R.} {\sc and} {\sc Sadri, F.} 2015.
\newblock Reactive computing as model generation.
\newblock {\em New Generation Computing\/}~{\em 33,\/}~1, 33--67.

\bibitem[\protect\citeauthoryear{Kowalski and Sadri}{Kowalski and
  Sadri}{2016}]{lpswithoutlp}
{\sc Kowalski, R.} {\sc and} {\sc Sadri, F.} 2016.
\newblock Programming in logic without logic programming.
\newblock {\em Theory and Practice of Logic Programming\/}~{\em 16,\/}~3,
  269--295.

\bibitem[\protect\citeauthoryear{Kowalski and Sergot}{Kowalski and
  Sergot}{1986}]{eventcalculus}
{\sc Kowalski, R.} {\sc and} {\sc Sergot, M.} 1986.
\newblock A logic-based calculus of events.
\newblock {\em New Generation Computing\/}~{\em 4,\/}~1, 67--95.

\bibitem[\protect\citeauthoryear{Lager and Wielemaker}{Lager and
  Wielemaker}{2014}]{DBLP:journals/tplp/LagerW14}
{\sc Lager, T.} {\sc and} {\sc Wielemaker, J.} 2014.
\newblock Pengines: Web logic programming made easy.
\newblock {\em {TPLP}\/}~{\em 14,\/}~4-5, 539--552.

\bibitem[\protect\citeauthoryear{{Nguembang Fadja} and Riguzzi}{{Nguembang
  Fadja} and Riguzzi}{2017}]{NguRig17-IMAKE-BC}
{\sc {Nguembang Fadja}, A.} {\sc and} {\sc Riguzzi, F.} 2017.
\newblock Probabilistic logic programming in action.
\newblock In {\em Towards Integrative Machine Learning and Knowledge
  Extraction}, {A.~Holzinger}, {R.~Goebel}, {M.~Ferri}, {and} {V.~Palade}, Eds.
  Springer, vol. 10344. Springer, Heidelberg, Germany.

\bibitem[\protect\citeauthoryear{Riguzzi}{Riguzzi}{2013}]{Rig13-FI-IJ}
{\sc Riguzzi, F.} 2013.
\newblock {MCINTYRE}: A {Monte Carlo} system for probabilistic logic
  programming.
\newblock {\em Fundamenta Informaticae\/}~{\em 124,\/}~4, 521--541.

\bibitem[\protect\citeauthoryear{Riguzzi, Bellodi, Lamma, Zese, and
  Cota}{Riguzzi et~al\mbox{.}}{2016}]{RigBelLam16-SPE-IJ}
{\sc Riguzzi, F.}, {\sc Bellodi, E.}, {\sc Lamma, E.}, {\sc Zese, R.}, {\sc
  and} {\sc Cota, G.} 2016.
\newblock Probabilistic logic programming on the web.
\newblock {\em Software: Practice and Experience\/}~{\em 46,\/}~10 (October),
  1381--1396.

\bibitem[\protect\citeauthoryear{Riguzzi and Swift}{Riguzzi and
  Swift}{2013}]{RigSwi13-TPLP-IJ}
{\sc Riguzzi, F.} {\sc and} {\sc Swift, T.} 2013.
\newblock Well\--definedness and efficient inference for probabilistic logic
  programming under the distribution semantics.
\newblock {\em Theory Pract. Log. Program.\/}~{\em 13,\/}~Special Issue 02 -
  25th Annual GULP Conference (March), 279--302.

\bibitem[\protect\citeauthoryear{Rossant}{Rossant}{2013}]{rossant2013learning}
{\sc Rossant, C.} 2013.
\newblock {\em Learning IPython for interactive computing and data
  visualization}.
\newblock Packt Publishing Ltd, Birmingham, UK.

\bibitem[\protect\citeauthoryear{Sato}{Sato}{1995}]{DBLP:conf/iclp/Sato95ijar}
{\sc Sato, T.} 1995.
\newblock A statistical learning method for logic programs with distribution
  semantics.
\newblock In {\em 12th International Conference on Logic Programming},
  {L.~Sterling}, Ed. {MIT} Press, Cambridge, MA, 715--729.

\bibitem[\protect\citeauthoryear{Sergot, Sadri, Kowalski, Kriwaczek, Hammond,
  and Cory}{Sergot et~al\mbox{.}}{1986}]{bna}
{\sc Sergot, M.~J.}, {\sc Sadri, F.}, {\sc Kowalski, R.~A.}, {\sc Kriwaczek,
  F.}, {\sc Hammond, P.}, {\sc and} {\sc Cory, H.~T.} 1986.
\newblock The british nationality act as a logic program.
\newblock {\em Communications of the ACM\/}~{\em 29,\/}~5, 370--386.

\bibitem[\protect\citeauthoryear{Srinivasan}{Srinivasan}{2007}]{aleph}
{\sc Srinivasan, A.} 2007.
\newblock Aleph.
\newblock \url{http://www.cs.ox.ac.uk/activities/machlearn/Aleph/aleph.html}.

\bibitem[\protect\citeauthoryear{Srinivasan, Muggleton, Sternberg, and
  King}{Srinivasan et~al\mbox{.}}{1996}]{DBLP:journals/ai/SrinivasanMSK96}
{\sc Srinivasan, A.}, {\sc Muggleton, S.}, {\sc Sternberg, M. J.~E.}, {\sc and}
  {\sc King, R.~D.} 1996.
\newblock Theories for mutagenicity: A study in first-order and feature-based
  induction.
\newblock {\em Artificial Intelligence\/}~{\em 85,\/}~1-2, 277--299.

\bibitem[\protect\citeauthoryear{Vennekens, Verbaeten, and
  Bruynooghe}{Vennekens et~al\mbox{.}}{2004}]{VenVer04-ICLP04-IC}
{\sc Vennekens, J.}, {\sc Verbaeten, S.}, {\sc and} {\sc Bruynooghe, M.} 2004.
\newblock {Logic Programs With Annotated Disjunctions}.
\newblock In {\em Logic Programming: 20th International Conference, ICLP 2004,
  Saint-Malo, France, September 6-10, 2004. Proceedings}, {B.~Demoen} {and}
  {V.~Lifschitz}, Eds. LNCS, vol. 3132. Springer Berlin Heidelberg, Berlin
  Heidelberg, Germany, 431--445.

\bibitem[\protect\citeauthoryear{Wielemaker and Anjewierden}{Wielemaker and
  Anjewierden}{2007}]{Wielemaker:2007c}
{\sc Wielemaker, J.} {\sc and} {\sc Anjewierden, A.} 2007.
\newblock {PlDoc}: {Wiki} style literate programming for {Prolog}.
\newblock In {\em Proceedings of the 17th Workshop on Logic-Based methods in
  Programming Environments}, {P.~Hill} {and} {W.~Vanhoof}, Eds. Cornell
  University Library, Ithaca, NY, 16--30.

\bibitem[\protect\citeauthoryear{Wielemaker, Huang, and van~der
  Meij}{Wielemaker et~al\mbox{.}}{2008}]{DBLP:journals/tplp/WielemakerHM08}
{\sc Wielemaker, J.}, {\sc Huang, Z.}, {\sc and} {\sc van~der Meij, L.} 2008.
\newblock Swi-prolog and the web.
\newblock {\em {TPLP}\/}~{\em 8,\/}~3, 363--392.

\bibitem[\protect\citeauthoryear{Worlfram}{Worlfram}{2016}]{SWblog}
{\sc Worlfram, S.} 2016.
\newblock How to teach computational thinking.
\newblock
  \url{http://blog.stephenwolfram.com/2016/09/how-to-teach-computational-thinking/}.

\end{thebibliography}

\end{document}